\documentclass[twocolumn]{aastex6}

\usepackage{epsfig}
\usepackage{graphicx}
\usepackage{longtable}
\usepackage{natbib}
\usepackage{subfigure}
\usepackage{xcolor}
\usepackage{amsmath}
\usepackage{multirow}
\usepackage{float}
\newcommand{\Msolar}{M$_{\odot}$}

\newcommand{\qcrit}{$q_\mathrm{crit}$}

\shorttitle{BSS Gaia DR2}
\shortauthors{Leiner et al.}

\begin{document}
\title{A Census of Blue Stragglers in Gaia DR2 Open Clusters as a Test of Population Synthesis and Mass Transfer Physics}

\author{Emily M. Leiner\altaffilmark{1}}
\author{Aaron Geller\altaffilmark{1,2}}

\email{emily.leiner@northwestern.edu}
\altaffiltext{1}{Center for Interdisciplinary Exploration and Research in Astrophysics (CIERA) and Department of Physics and Astronomy, Northwestern University, 1800 Sherman Ave., Evanston, IL 60201, USA}
\altaffiltext{2}{Adler Planetarium, Department of Astronomy, 1300 S. Lake Shore Drive, Chicago, IL 60605, USA}

\begin{abstract}
We use photometry and proper motions from Gaia DR2 to determine the blue straggler star (BSS) populations of 16 old (1-10 Gyr), nearby ($d< 3500$ pc) open clusters. We find that the fractional number of BSS compared to RGB stars increases with age, starting near zero at 1 Gyr and  flattening to $\sim 0.35$ by 4 Gyr. Fitting stellar evolutionary tracks to these BSS, we find that their mass distribution peaks at a few tenths of a solar mass above the main-sequence turnoff. BSS more than 0.5 \Msolar\ above the turnoff make up only $\sim25$\% of the sample, and BSS more than 1.0 \Msolar\ above the turnoff are rare. We compare this to Compact Object Synthesis and Monte Carlo Investigation Code (\texttt{COSMIC}) population synthesis models of BSS formed via mass transfer. We find that standard population synthesis assumptions dramatically under-produce the number of BSS in old open clusters. We also find that these models overproduce high mass BSS relative to lower mass BSS. Expected numbers of BSS formed through dynamics do not fully account for this discrepancy. We conclude that in order to explain the observed BSS populations from Roche lobe oveflow, mass-transfer from giant donors must be more stable than assumed in canonical mass-transfer prescriptions, and including non-conservative mass transfer is important in producing realistic BSS masses. Even with these modifications, it is difficult to achieve the large number of BSS observed in the oldest open clusters. We discuss some additional physics that may explain the large number of observed blue stragglers among old stellar populations. 
\end{abstract}

\section{Introduction}

One discovery of the first cluster color-magnitude diagrams was that members stars often could be found brighter and bluer than the main-sequence turnoff of the cluster. These blue straggler stars (BSS) should have evolved into giants millions or billions of years ago. Several explanations have been proposed to explain how blue stragglers may form. They may result from stellar collisions during dynamical encounters \citep{Leonard1989}; mergers of close binary systems, perhaps the inner binaries in triple systems driven to merger by Kozai cycles \citep{Perets2009}; or the result of mass-transfer from a binary companion \citep{McCrea1964}. It is likely that all these mechanisms contribute to the blue straggler population to varying degrees. 

Blue stragglers are found in both open and globular clusters. Perhaps the most detailed study of a blue straggler population to-date has been in the old open cluster NGC 188 (6 Gyr).  80\% of the blue stragglers in the cluster are spectroscopic binaries \citep{MathieuNature2009, mathieu2015, Gosnell2015}. \citet{Geller2011} found that the statistical mass distribution of the companions to these blue stragglers peaks near 0.5 $M_\odot$, suggestive of white dwarf secondary stars.  Hubble Space Telescope UV photometry later confirmed that 7 of the 20 blue stragglers in this cluster host hot white dwarf companions that are less than 400 Myr old, pointing to recent mass transfer from a red giant branch (RGB) or asymptotic giant branch (AGB) companion \citep{Gosnell2014, Gosnell2015}. The authors conclude that 2/3 of the blue straggler population in the cluster likely formed through this channel, since some of the blue stragglers would likely have older, fainter white dwarfs below the detection limit of the study. 

Such detailed observations are not available in other clusters, but other studies also point to mass-transfer as the likely origin for most blue stragglers in old clusters. In globular clusters, the lack of correlation between cluster density and blue straggler formation suggests that internal binary evolution, including mass transfer, (rather than dynamical collisions) may be the dominant formation channel there \citep{Knigge2009}. The significant population of blue stragglers and other AFG-type post-mass-transfer binaries in the field also indicates mass transfer is likely a common formation channel \citep{Murphy2018, Escorza2019, Carney2001, Carney2005}. These populations therefore offer a snapshot of post-mass-transfer outcomes in solar-type binaries. Yet detailed studies of blue straggler populations, especially across large populations including different cluster ages, densities, and metallicities, are still scarce. The most comprehensive catalog of blue stragglers in open clusters is given in \citet{Ahumada1995} and updated in \citet{Ahumada2007}, although the data quality is inhomogenous and many of the clusters included have limited astrometric membership information.  

The release of Gaia DR2, proper-motions and parallaxes to 1.3 billions stars allows for kinematic memberships for any nearby star cluster, enabling a much improved census of their blue straggler populations. Here we use Gaia DR2 to select members of a sample of 16 old ($> 1.0$ Gyr), nearby (d $\leq 3500$ kpc) rich open clusters.

We use this sample to study the blue straggler populations of these clusters, comparing them to predictions from binary population synthesis models. This sample can provide important constraints on the formation frequencies of blue straggler stars and on the physics of mass transfer in low mass (M $\leq$ 2.0 \Msolar) stars. This mass transfer physics is important not only for blue straggler stars, but for many other astrophysical objects that result from binary evolution in low-mass stellar systems (e.g. double white dwarfs, Type Ia supernovae and other transients resulting from white dwarf mergers, X-ray binaries, etc.)

In this paper we first present our technique for selecting clusters and determining membership (Section~\ref{sec:sample}), then present analysis of the number of blue stragglers in each cluster and search for trends in these results (Section~\ref{sec:NBSS}). We compare these observations to ~\texttt{COSMIC} \citep{Breivik2020} population synthesis models to determine how well these models reproduce observed cluster blue straggler populations (Section~\ref{sec:BSE}). Finally, we discuss discrepancies between our observations and models, and consider some of the mass-transfer physics that may need to be improved in population synthesis in order to obtain more realistic results (Sections~\ref{sec:results}~and~\ref{section:discussion}). In Section~\ref{sec:summary} we provide our summary and conclusions.

\section{Cluster Sample}\label{sec:sample}
\subsection{Initial Selection}
We start by compiling a list of open clusters that is as complete as possible by combining multiple catalogs in the literature. Specifically, we use the Milky Way Star Clusters Catalog \citep[MWSC][]{MWSC1, MWSC2, MWSC3, MWSC4}, \citet{Lynga}, \citet{Piskunov2008}, \citet{Salaris2004}, \citet{vandenBergh2006}, \cite{CantatGaudin2018}, and we also download and include all the open clusters from WEBDA\footnote{https://webda.physics.muni.cz/}.   From this catalog we select the clusters that have masses $>200M_\odot$, distances $<3.5$kpc and ages $>1$ Gyr. For clusters that have multiple mass, distance, and/or age values from the references given above, we simply take the mean value. This results in a list of 39 open clusters. Cluster parameters in Table~\ref{tab:clusters} are drawn from this catalog unless otherwise specified. 

\subsection{Gaia Memberships} 
\label{section:memberships}
We perform a membership selection for all 39 clusters in our sample using using Gaia astrometric measurements as follows. 

First, we use literature values of cluster RA and Dec. We select from the Gaia catalog all stars with $g< 18.0$ within 0.1 arcmin of this cluster central coordinate. This small radius limits our selection to stars in the central regions of the cluster that are very likely to be true cluster members. From this initial sample, we fit a Gaussian to the distribution of distances \citep{BailerJones2018}, to the distribution of the Dec. proper motion measurements, and to the distributions of the RA proper motion measurements. These fits give us an initial estimate of the distance to the cluster and the proper motion of the cluster, which we use as input parameters for the fits below.

We then select all stars within an angular distance corresponding to 6 pc, 8 pc, 12 pc, 16 pc, 20 pc, or 24 pc of this central coordinate, assuming the cluster distance computed above. These apertures allow for a reasonable range of open cluster sizes. In some cases, we find that within all apertures the cluster members are not clearly distinguishable from the field in our proper motion distribution. We find 8 such clusters and remove them from our sample. In one case (the cluster Dias 4) we find that there are actually two clusters at this location, and remove this from our sample as well. 

For each remaining cluster, we find the first aperture in which the number of RGB members of the cluster increases by $< 5\%$ from the next smaller aperture size, and use this as the aperture for the rest of our analysis. At this 5\% level we are adding 0-3 RGB stars to most clusters, depending on cluster size. At this level we are no longer adding many cluster stars, but will continuing to add field contamination as we increase the aperture.  This threshold therefore allows us to select an aperture that will contain most cluster members, while ensuring we don't select an unnecessarily large aperture that could increase field contamination. We find that this is typically $\approx 3-5$ times the half mass radius of a cluster. We show the aperture size we select (r) and the mean half-mass radius (r$_\text{hm}$) from our catalog compilation in Table~\ref{tab:clusters}.  

The Gaia magnitude limit is G= 21.0, but incompleteness, astrometric errors, and photometric errors become larger at the faint end. We therefore select all stars in the Gaia DR2 database brighter than G= 18. This magnitude limit is inclusive of solar-type main sequence stars in old clusters out to 3.5 kpc, but eliminates faint stars where the astrometry may be poor. We flag sources with non-zero values for the parameter ~\texttt{astrometric\_excess\_noise}. These sources have larger than expected errors in the Gaia astrometric solution, possibly due to the presence of a binary companion that perturbs their motion. The errors on the parallax and proper motion of these stars may be larger than typical, and therefore we exclude these sources from the fits described below. We do include these sources in our membership determinations, though their memberships status is less certain. 

We then fit two 2D Gaussians to the proper motion distribution. One Gaussian fits the field proper motion distribution, and the other fits the cluster proper motion distribution. For each source, we then determine a cluster proper motion membership probability by dividing the value of the cluster Gaussian by the sum of the cluster and field Gaussians at the proper motion of that source. 

For each proper-motion member, we find the distance determined by \citet{BailerJones2018}. We fit a Gaussian function to the distribution of these distances. We determine the cluster distance by taking the mean of this Gaussian. We note that there may be small offsets in these distances due to complex Gaia systematics that have not been well quantified. These errors, however, are expected to be small and do not significantly impact our membership analysis. 

For the color-magnitude diagrams presented below, we use membership cutoffs of $P_\mathrm{PM} > 50\%$ to qualify as a member. We also require that the \citet{BailerJones2018} distance to the star is $< 3 \sigma$ from the cluster mean to exclude any clear field interlopers. We do not use any radial-velocity (RV) membership information, as Gaia RVs are only available for a small subset of bright stars in our sample. 

\begin{figure}
    \centering
    \includegraphics[width=.95\linewidth]{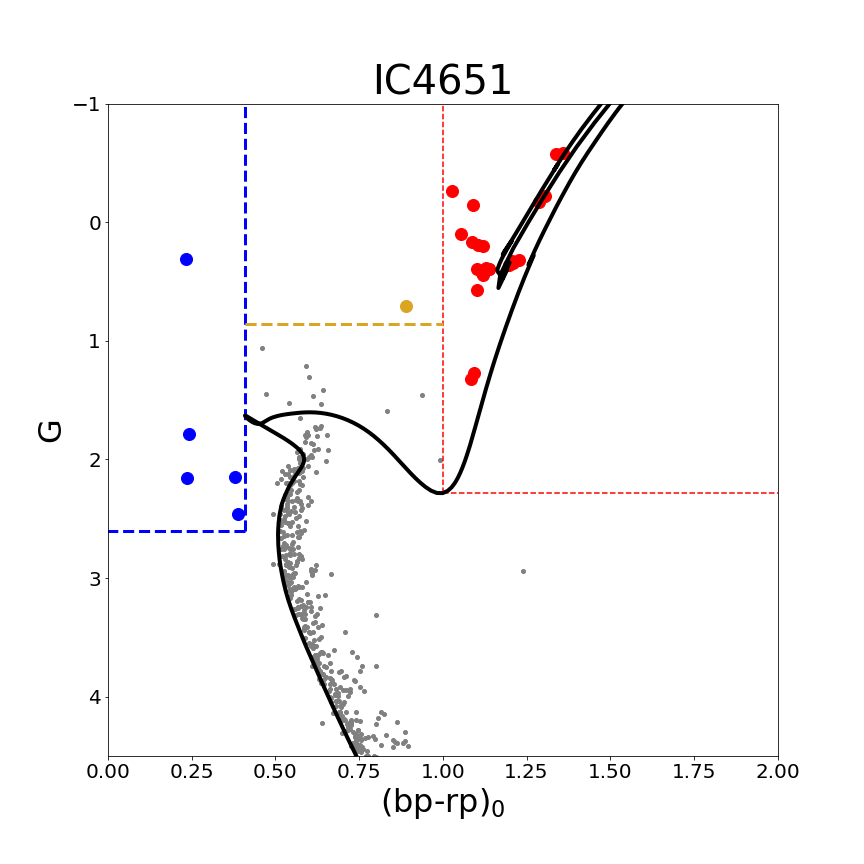}
    \caption{A color-magnitude diagram of IC4651 showing Gaia members of the cluster (gray points). Lines separate the RGB stars (red), blue straggler stars (blue) and yellow straggler stars (between blue and red, above the yellow line). In black we show a 1.4 Gyr MIST isochrone. We show this figure as an example; CMDs for the rest of the clusters in our sample are in the Appendix(Figure~\ref{fig:allclustercmds}).}\label{fig:examplecmd}
\end{figure}

We also experiment with using proper motion membership cuts of 20\% and 80\%. In most cases these different cuts do not significantly change the blue straggler counts in a cluster. In two cases where the field and cluster proper-motion distributions are not well separated, different cuts result in blue straggler populations that vary by more than $1 \sigma$ from the value using a $50\%$ cut. In these two cases the blue straggler memberships are more uncertain, and so we exclude these clusters from our analysis.

\subsection{Reddening Corrections}
\label{section:reddening}
In clusters with low reddening (E[bp-rp] $<$ 0.3), we simply adopt reddening measurements from the literature, converting to the Gaia pass bands using \citet{Jordi2010}. We list these global reddening values in Table 1. We compare these literature values also to the mean value found for the cluster using a 3D dust map \citep{Green2019} where available (Dec. $>$ -30 degrees). This map is an improvement over typical 2D dustmaps, which provide integrated values of extinction along a given line-of-sight, but do not provide distance-dependent corrections. \citet{Green2019} uses far infrared emission to map the 2D dust distribution across the sky, and then incorporates stellar photometry and parallax measurements from Gaia and other surveys, using the observed reddening of these stars to infer the distance distribution of the absorbing dust. We find reddening values from this dustmap to be in reasonable agreement with previous literature estimates. 

In three clusters with considerable reddening, we compute star-by-star reddening corrections using this dust map \citep{Green2019}. Using these corrections yields narrower features in the cluster CMDs by correcting for the effects of substantial differential reddening. We elect to use these corrections for our analysis of NGC 7789, NGC 6939, and Berkeley 17. For these clusters we use star-by-star reddening corrections in our CMDs, and we compute the mean reddening measurement of all cluster members and report this mean reddening value in Table 1. Despite its high reddening, we find that the star-by-star reddening corrections in Collinder 110 broaden the CMD features considerably. Therefore, we adopt the mean cluster value as the global cluster reddening for Collinder 110.

In some cases, particularly for distant ($> 3$ kpc) and highly extincted clusters (E(bp-rp) $> 0.7$), the dust map is not available (Dec$ < -30$ degrees) or does not do an adequate job in correcting for differential extinction. A large spread in color of the main sequence and giant branch remains, which creates uncertainty in selecting blue straggler candidates near the main sequence turnoff. We exclude 9 clusters with very high reddening from our blue straggler analysis for this reason.

\subsection{Sparse Clusters}

Finally, we remove 2 clusters from our initial list that are sparsely populated with fewer than 10 RGB stars. These clusters likely have over-estimated masses in the literature, and thus are not rich enough to rely on isochrone-derived ages or confidently identify blue straggler populations. 

After this membership and reddening analysis, we are left with 16 clusters from our sample with high-quality memberships and color-magnitude diagrams. These 16 clusters are the basis of our analysis below. Properties for these clusters can be found in Table ~\ref{tab:clusters}. 

\begin{table*}
    \centering
    \begin{tabular}{l|c|c|r|c|c|c|c|c|c|l}
        \hline
         Cluster & Distance & [Fe/H] & Age &  $E$($bp-rp$) & r$_{hm}$ & r & N$_\text{BSS}$ & $N_\text{YSS}$ & N$_\text{RGB}$ \\
         & (pc) & & & & (pc) & (pc)& & & \\
         \hline
         \hline
         NGC 752 & 440 $\pm$ 20 & 0.0 & 1.4 Gyr & 0.04 & 4.2 & 16 & 1& 0 & 15 \\
         NGC 7789& 2080 $\pm$ 190 & 0.0 &  1.4 Gyr & 0.38 & 6.4 & 20 & 10 & 0 & 165 \\
         Collinder 110 & 2200 $\pm$ 300 & 0.0 & 1.6 Gyr  & 0.50 & 5.4 & 20 & 8 & 0 & 74\\
         NGC 6939 & 1850 $\pm$ 140 & 0.0 & 1.6 Gyr & 0.44 & 2.3 & 16 & 4 & 0 & 44\\

         IC 4651 & 920 $\pm$ 50 & 0.0 & 1.8 Gyr & 0.18 & 1.7 & 8& 4 & 1 & 24\\
 
         NGC 2506 & 2900 $\pm$ 500 & --0.25 & 2.0 Gyr & 0.08 & 5.0 & 16 & 9 & 0 & 72 \\
         NGC 6819 & 2600 $\pm$ 400 & +0.25 & 2.5 Gyr & 0.10 & 2.4 & 16 & 17 & 0 & 76 \\
         Ruprecht 147 & 300 $\pm$ 10 & 0.0 & 2.8 Gyr &  0.08 & 2.2 & 8 & 3 & 0 & 10\\
         Ruprecht 171 & 1530 $\pm$ 150 & +0.4  & 2.8 Gyr & 0.26 & 1.8 & 8 & 5 & 0  &33\\
         NGC 6253 & 1670 $\pm$ 160 & +0.4 & 3.2 Gyr & 0.28 & 1.6 & 8 & 16 & 1  & 55\\
         Berkeley 98 & 3400 $\pm$ 160 & 0.0 & 3.5 Gyr & 0.22 & 2.3 & 8 & 5 & 1  & 21\\
         M67 & 840 $\pm$ 60&0.0 & 4.0 Gyr & 0.05 & 2.5 & 12 & 16 & 0 & 43\\
         NGC 2243 & 3500 $\pm$ 700 & --0.5& 4.5 Gyr & 0.06 & 2.6 & 12 & 14 & 0 & 34\\
         NGC 188 & 1830 $\pm$ 170 & 0.0 & 6.3 Gyr & 0.12 & 4.6 & 16 & 16  & 0 & 60 \\
         Berkeley 39 & 3200 $\pm$ 800 & --0.25 & 8.0 Gyr & 0.10 & 2.1 & 12 & 24& 4 & 70\\
         Berkeley 17 & 2800 $\pm$ 750 & --0.25 & 10 Gyr & 0.62 & 2.7 & 12 & 20 & 2 & 59\\
         \hline
    \end{tabular}
    \caption{Parameters of open clusters in our sample adopted for this study, including the mean Gaia distance of all members \citep{BailerJones2018}, isochrone age and metallicity \citep{Dotter2016}, average cluster reddening, cluster half-mass radius (r$_\mathrm{hm}$), and radius for membership selection (r) . We also give blue straggler, yellow straggler and RGB counts.}
    \label{tab:clusters}
\end{table*}

\subsection{Isochrones}
We determine the appropriate MIST isochrone \citep{Dotter2016} for each cluster as follows. First, we determine the metallicity for the cluster from literature sources (provided in Table~\ref{tab:clusters}). In most cases the metallicity is near solar, and we adopt solar metallicity isochrones. We adopt reddening and extinction parameters listed in Table 1 (see Section~\ref{section:reddening}). We use a distance to the cluster of the mean of all distances to cluster members (Section~\ref{section:memberships}). Based on these parameters, we pick the isochrone by eye from a grid of MIST models with a spacing of log($\delta t= 0.05$). We find that catalog age of the cluster agrees well with our selected age in all cases, and list the isochrone ages in Table 1. We stress here that we do not intend these as improved Gaia determination of cluster ages, but rather to verify previous age measurements are reasonable and that the isochrone fits the main-sequence turnoff adequately, which is important for our blue straggler selection criteria (described below).

\section{Blue Straggler Populations in Gaia Clusters}\label{sec:NBSS} 

\subsection{Cluster CMDs} 
We show an example color-magnitude diagram (CMD) for IC 4651 in Figure~\ref{fig:examplecmd}. CMDs for our entire sample can be found in the Appendix ( Figure~\ref{fig:allclustercmds}). The CMDs are corrected assuming the distance and reddening values given in Table~\ref{tab:clusters}. For each cluster, we also show a MIST isochrone using the age and metallicity listed in Table~\ref{tab:clusters}. We plot all Gaia members with gray points, and highlight blue stragglers (blue points), yellow stragglers (yellow points) and red giants (red points). We discuss how we select these stars below. 

\subsection{Defining the Blue Straggler Domain}
There is substantial variation in the definition of blue straggler stars between different observational works \citep{Leigh2011b}. 

For this work, we define blue stragglers as being bluer than the bluest point on the cluster isochrone. If the cluster has a blue hook, we set this limit at the bluest point in the isochrone.  This excludes the region slightly blue of the main sequence just below the blue hook; some true blue stragglers may reside in this area, but because normal photometric binaries may fall in this region and because the exact shape of the hook is sensitive to detailed properties of stellar models such as overshooting, we cannot confidently classify these stars as blue stragglers. If the cluster does not have a blue hook, we require that the blue straggler is at least 0.03 bluer than the bluest point on the isochrone, which allows for some photometric scatter among normal main sequence stars.  For all clusters, we also require that the blue straggler is no more than 1 magnitude below the cluster turnoff. This magnitude cut excludes white dwarf-main sequence binaries, which sometimes are present at high temperatures and low luminosities well below the turnoff. It also allows us to consider similar CMD regions regardless of cluster distance, as our magnitude limit of g= 18 is only about 1 magnitude below the turnoff in our oldest and most distant clusters. Visual examination shows this magnitude cut rarely excludes possible blue stragglers from our analysis. We mark this blue straggler region with the blue dashed lines in Figure \ref{fig:examplecmd}. 

Some stars, called ``yellow stragglers" or ``evolved blue stragglers" have been found in clusters \citep{Leiner2016, Landsman1997}. These stars are significantly brighter than the main sequence, but redder than blue stragglers, falling in between the blue straggler region and red giant branch. These stars are thought to be blue stragglers that have evolved into giant or subgiant stars. We define the yellow straggler region to be more than 0.75 magnitudes above the brightest point on the subgiant branch of the isochrone (the brightest magnitude expected for the combined light of an equal mass binary at the cluster turnoff), with a color redder than the defined blue straggler region and bluer than the base of the RGB (see Figure~\ref{fig:allclustercmds}). These stars are rare, and in the remainder of our analysis we combine the blue straggler and yellow straggler populations to determine $N_\mathrm{BSS}$. 

We classify stars that are brighter and redder than the base of the RGB on our isochrone as standard RGB stars. 

Using this approach we count the total number of red giants, blue stragglers and yellow stragglers within the color-magnitude diagrams of our 16 open clusters. We report the number of such stars in Table 1, along with the aperture we used to determine these star counts. We plot the observed ratio of $\frac{N_\text{BSS}}{N_{RGB}}$ in the left panel of Figure ~\ref{fig:BSS_RGB_ratio}. We note that the results of these star counts are not very sensitive to the aperture adopted, and adjusting the aperture size up or down one step results in minimal changes to our results (i.e. $<10\%$ changes in star counts; differences in $\frac{N_\text{BSS}}{N_{RGB}} < 3$\%)



\subsection{Combined CMDs}

\begin{figure*}
     \centering
     \includegraphics[width=.3\linewidth]{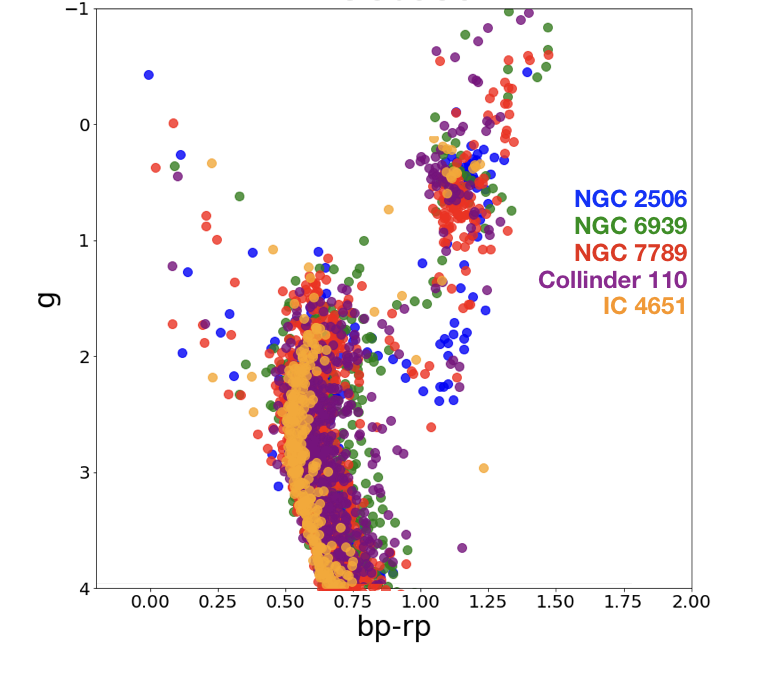}
     \includegraphics[width=.3\linewidth]{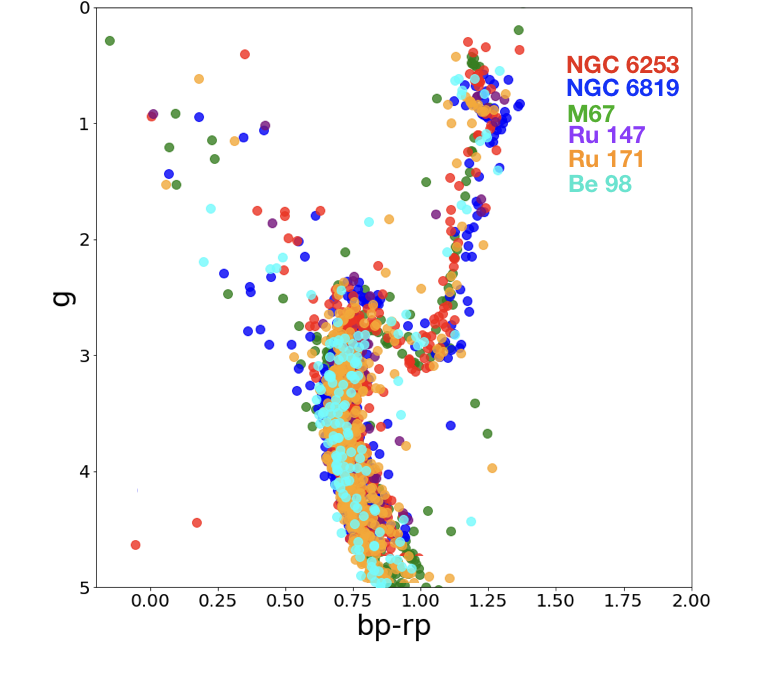}
     \includegraphics[width=.3\linewidth]{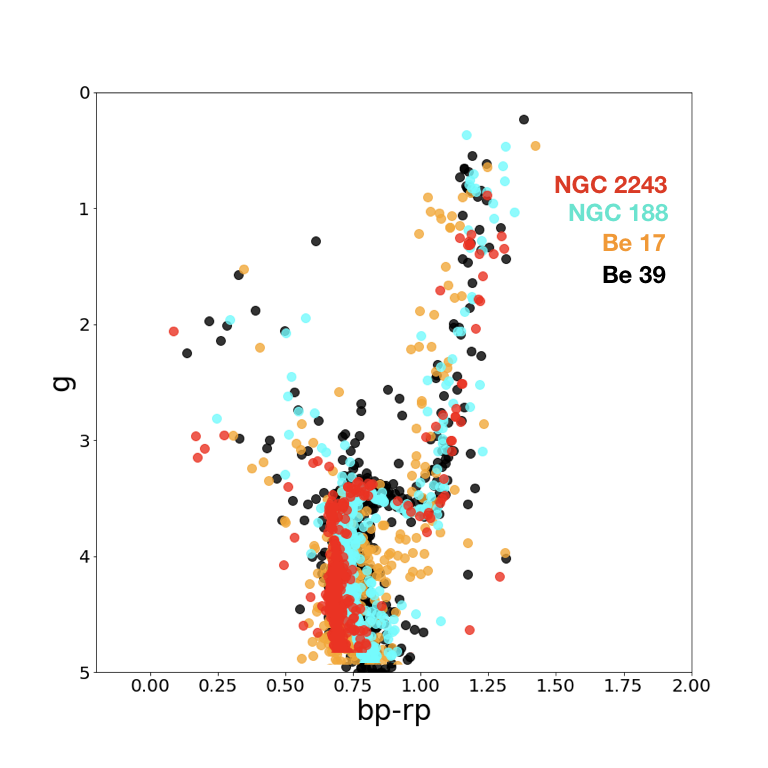}
     \caption{We show a combined CMDs of the youngest open clusters ($<2$ Gyr; left panel), middle aged open cluster (2-4 Gyr; middle panel), and oldest open clusters ($>$ 4 Gyr; right panel) in our sample. Each shows absolute g magnitude versus de-reddened bp-rp color. For each cluster, we shift the color and magnitude by a small amount to align the main sequence turnoffs of all clusters in the plot.}
     \label{fig:combined_cmd}

 \end{figure*}

We show in Figure~\ref{fig:combined_cmd} combined color-magnitude diagrams made by plotting the CMDs of every cluster in our sample, shifting the photometry by a small amount $\delta G$ and $\delta(bp-rp)$ such that the main sequence turnoffs of each cluster align. This allows us to compare the positions of the blue straggler stars relative to the main sequence turnoff in a much larger sample than any one cluster alone. For these plots, we split our sample into 3 bins by age: 1-2 Gyr, 2-4 Gyr, and more than 4 Gyr. This allows us to compare clusters with similar turnoff masses, and splits our sample into 3 groups containing 4-5 clusters each. For clarity, we exclude clusters that have 1 or fewer blue stragglers.

We see in Figure~\ref{fig:combined_cmd} that for intermediate age clusters there appear to be two clumps of blue stragglers, one brighter clump about 1.5 magnitudes above the turnoff, and one fainter clump stretching from the MSTO to about 1 magnitude above. A less prominent, narrower gap may also be present in the old clusters ($> 4$ Gyr). For the youngest clusters (1-2 Gyr) a clear gap is not visible. For the total sample of old and intermediate aged clusters, 30\% of the blue stragglers are found in the bright blue straggler region, and 70\% are in the fainter region. 

\subsection{Blue Straggler Masses}
\begin{figure*}
    \centering
    \includegraphics[width=1.0\linewidth]{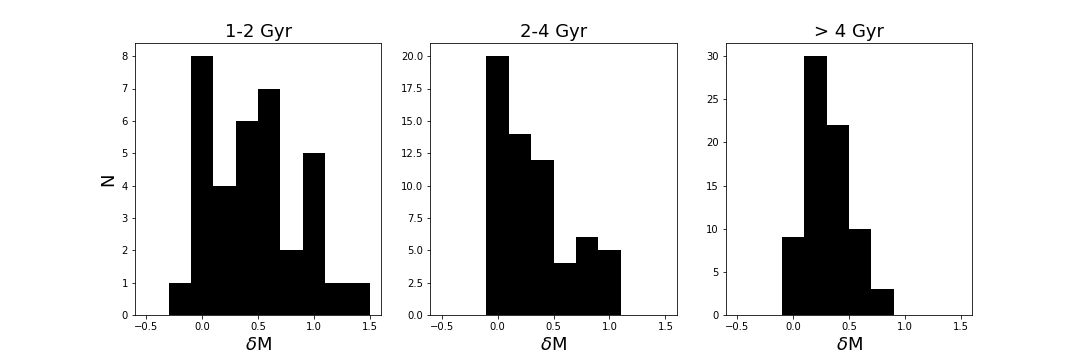}
    \caption{A histogram of the number of blue stragglers in our cluster sample binned by the difference between the blue straggler mass and the turnoff mass ($\delta M$). We show from left to right the sample of young clusters (1-2 Gyr), intermediate aged clusters (2-4 Gyr), and old clusters ($>$4 Gyr). For context, the turnoff mass in a 1-2 Gyr cluster would be 2.0-1.6 \Msolar\, 1.6-1.3 \Msolar\ in the 2-4 Gyr clusters, and 1.3-1.0 \Msolar\ in a 4-10 Gyr cluster.}
    \label{fig:masshist}
\end{figure*}

We determine masses for each blue straggler by fitting MIST evolutionary tracks \citep{Dotter2016} to the CMD positions of the blue stragglers in our clusters using the metallicity and cluster parameters given in Table~\ref{tab:clusters}. We show histograms of the derived blue straggler masses in Figure~\ref{fig:masshist}, expressing them in terms of the difference between the turnoff mass of the cluster and the blue straggler mass ($\delta M= M_\mathrm{BSS} - M_\mathrm{MSTO}$). 

Clusters of all ages show a distribution skewed towards lower mass blue stragglers, with a peak 0.3 \Msolar\ above the turnoff among the entire sample of 171 blue stragglers. Only 5 stars in the whole sample have masses  $\gtrsim 1.0$ \Msolar\ above the turnoff, and no blue stragglers are more than 1.5 \Msolar\ above the turnoff. For context, we note that the maximum $\delta M$ expected for conservative AGB mass transfer onto an MSTO star would lead to a BSS with $\delta M\eqsim 1.5$ in the young clusters, $\delta M\eqsim 1.0$ in the intermediate aged clusters, and $\delta M \eqsim 0.8$ in the oldest clusters. The most massive blue stragglers in these distributions, then, are roughly compatible with these limits. We find no blue stragglers so massive they must have formed from mergers or collisions of two or more normal cluster stars. We note also, that most BSS are significantly less massive than this limit. If they formed from mass transfer, this would require progenitor accretors with masses significantly below the turnoff and/or quite non-conservative mass transfer. We return to this finding, as well as the overall shape of the blue straggler mass distribution, in Section~\ref{section:discussion}.

\subsection{Blue Straggler Fractions} 
Given that more blue stragglers should be expected in larger clusters, we scale the number of blue stragglers observed in the cluster by the number of RGB stars in the cluster. We use this scaling because RGB stars are easy to count in the cluster, and because the RGB population traces the number of stars near the main-sequence turnoff. This is likely to be the progenitor population of the blue straggler stars formed through mass-transfer. We show this ratio as a function of cluster age in Figure ~\ref{fig:BSS_RGB_ratio} (black points).
We find that the blue straggler fraction increases with cluster age from 1-4 Gyr, and then plateaus at a near constant $\frac{N_\mathrm{BSS}}{N_\mathrm{RGB}} \approx 0.35$ at later ages. We find no trends with cluster distance or reddening. Most of our sample consists of clusters of near-solar metallicity, and we do not have enough high and low metallicity clusters to determine whether there are differences in blue straggler fraction in clusters of different compositions. 

\citet{Ahumada1995} also found in their catalog that old clusters have more blue stragglers than young clusters. With our result, we confirm that this trend holds with the more accurate photometry, astrometric memberships, and cluster parameters enabled by Gaia, and provide a more well-sampled picture of the evolution with age for clusters older than 1 Gyr. 

In the next section, we investigate whether population synthesis models can produce the blue straggler populations observed or the trend seen with increasing blue stragglers in older clusters.

\section{Population Synthesis with~\texttt{COSMIC}}\label{sec:BSE}
We use the population synthesis code~\texttt{COSMIC} \citep{Breivik2020} \footnote{https://github.com/COSMIC-PopSynth/COSMIC} to model cluster blue straggler populations and determine if these models reproduce the trend we see between blue straggler population and cluster age.~\texttt{COSMIC} implements the population synthesis code Binary Stellar Evolution (~\texttt{BSE} \citealt{Hurley2002}) inside an easy-to-use ~\texttt{python} wrapper, and includes some updates to the binary evolution physics. 

\subsection{Prescriptions for $q_\text{crit}$}
\begin{figure*}
    \centering
    \includegraphics[width= .95\linewidth]{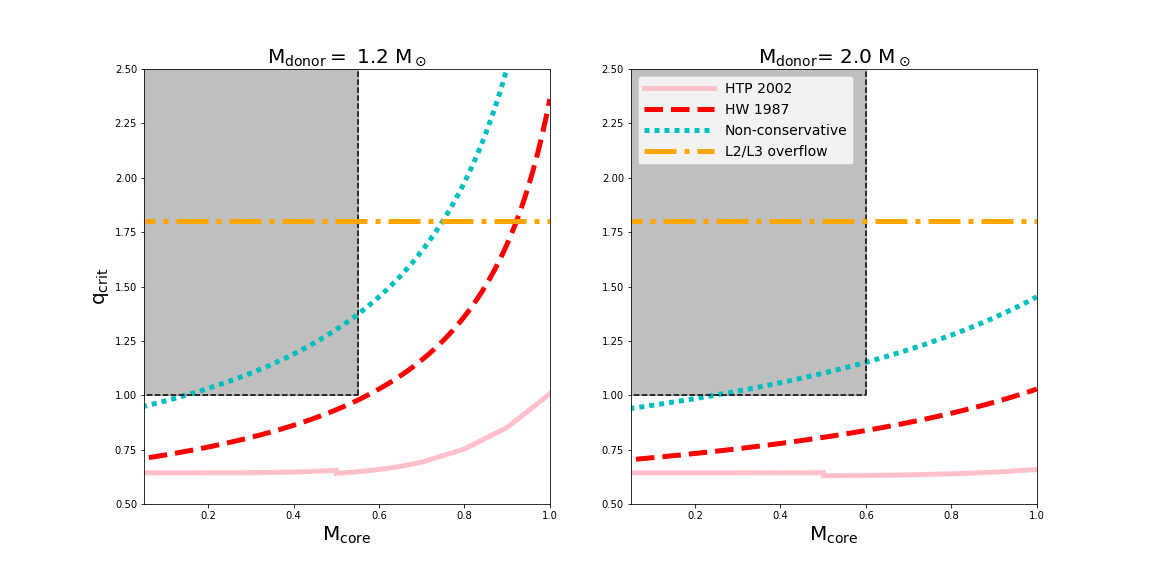}
    \caption{A plot of the critical mass ratio for mass-transfer stability ($q_\mathrm{crit}$) as a function of the donor core mass. We assume a donor mass of 1.2 \Msolar on the left and 2.0 \Msolar on the right. The colored lines represent different $q_\mathrm{crit}$ prescriptions. The gray shaded region shows the region where the donor is more massive than the accretor, and the core mass of the donor is consistent with expectations for a low mass RGB or AGB star. This is the region where stable mass transfer would have to occur in order to produce blue stragglers.}
    \label{fig:MT_prescriptions}
\end{figure*}

We focus our investigation on the impact of the choice of $q_\text{crit}$, the mass ratio required for mass-transfer stability. Population synthesis models do not compute full stellar structure models to determine the outcome of mass transfer in a binary system. Instead, these models use fitting formulae to predict whether mass transfer will proceed stably-- resulting in significant mass transfer and possibly producing a blue straggler-- or unstably-- resulting in a common envelope (CE), little mass accretion by the companion, orbital shrinkage, and possibly a merger of the binary system. For giant donor stars, this prescription typically depends on the mass fraction of the donor star's core, $M_\mathrm{c1}/M_1$, and the mass ratio of the binary system ($q=\frac{M_\mathrm{1}}{M_\mathrm{2}}$). 

Because many binaries will interact as the donor star expands into an RGB or AGB star, the critical mass ratio for mass transfer from a giant donor is essential to understanding the outcomes of mass transfer. This mass ratio for stability has been the subject of much debate in the literature, and varying prescriptions have been implemented across different population synthesis codes. There are a large range of prescriptions in use that can yield dramatically different mass-transfer outcomes. 

We show in Figure~\ref{fig:MT_prescriptions} four examples of stability criteria, all of which we test in our~\texttt{COSMIC} models. In this example, we show these criteria for both a 1.2 \Msolar\ giant donor star on the left, and a 2.0 \Msolar\ giant donor on the right (though in our population synthesis models, we apply these criteria at the modeled mass of the star). These masses are at the upper and lower range of donor stars we would expect in our cluster sample given our age range of $\sim1-10$ Gyr.  We implement these criteria for giant star donors only (in ~\texttt{BSE}, ~\texttt{kstar} types 3, 4, 5, 6), and use the default \citet{Hurley2002} prescription for all other stellar types. We describe each of these four models briefly below. 

\subsubsection{\citet{Hurley2002}}
In ~\texttt{BSE}, the default option for $q_\mathrm{crit}$ for mass-transferring binaries with a giant donor star is given by the equation: 

\begin{equation}\label{eq:HTP2002}
    q_\mathrm{crit}= \left[1.67-x+2\left(\frac{M_\mathrm{c1}}{M_1}\right)^5\right]/2.13
\end{equation}

Here $M_\mathrm{C1}$ is the core mass of the donor star, $M_1$ is the total mass of the donor star, and $x$ is the helium mass fraction of the star. For mass transfer to proceed stably in population synthesis models that use this presciption, $\frac{M_\mathrm{donor}}{M_\mathrm{accretor}} > q_\mathrm{crit}$. This is Equation 57 in \citet{Hurley2002} and derived by fitting results from detailed stellar evolution calculations. Note that this model assumes mass transfer is conservative (i.e., all the material lost by the donor is accreted by the companion and no angular momentum is lost from the system). This equation is shown in Figure~\ref{fig:MT_prescriptions} with the pink line. 

\subsubsection{\citet{Hjellming1987}}
An alternative stability criterion is given by \citep{Hjellming1987}. This equation is determined analytically using condensed polytropes as stellar models. Like the \citet{Hurley2002} expression, this assumes fully conservative mass transfer. This model is shown with the red line in Figure~\ref{fig:MT_prescriptions}: 

\begin{equation}
 q_\mathrm{crit}= 0.362+\left[3\left(1-\frac{M_\mathrm{c1}}{M_1}\right)\right]^{-1}
\end{equation}

with variables defined as in Equation~\ref{eq:HTP2002}. 

\subsubsection{Non-conservative Mass Transfer}\label{section:nonconservativeMT}
\citet{Woods2011} demonstrate that stability criteria like the \citet{Hurley2002} and \citet{Hjellming1987} prescriptions underestimate the stability of mass transfer because of the assumption that mass transfer will be conservative. \citet{Woods2011} re-derive the stability predictions including a mass transfer efficiency parameter, $\beta$, that represents the fraction of material lost by the donor that is accreted by the companion. When $\beta$ is small (i.e. mass transfer is highly non-conservative), mass transfer can proceed stably for higher mass ratio systems.

Here we use a Hjellming-like model that approximates the non-conservative stability criteria of \citet{Woods2011} assuming $\beta=20$\% over the relevant range of core masses for low-mass giants (M$_\mathrm{core} < 0.6$). Specifically, the equation we implement is: 
\begin{equation}
q_\mathrm{crit}=0.4+\left[1.9\left(1-\frac{M_\mathrm{c1}}{M_1}\right)\right]^{-1}
\end{equation}

This stability criterion is shown with the cyan line in Figure~\ref{fig:MT_prescriptions}. We note that while we use this relation that assumes non-conservative mass transfer to determine whether a binary will evolve to common envelope or stable mass transfer, ~\texttt{BSE} determines mass transfer efficiency independently from stability calculations by comparing the Kelvin-Helmholtz timescale of the accretor to the mass-transfer time scale (\citealt{Hurley2002}). That is, though we assume 20\% efficiency in this relation, ~\texttt{BSE} will still assume conservative mass transfer unless the calculated mass loss rate exceeds a thermal timescale of the accretor. A fully consistent treatment of mass transfer with this model would couple the calculation of the mass-transfer stability with a determination of mass-transfer efficiency, but this is beyond the scope of this paper. For now, we adopt this criterion simply to provide an upper limit on how many blue stragglers may be created if allowing for non-conservative mass transfer with a Hjellming-like law. 

\subsubsection{L2/L3 Overflow}
\citet{Pavlovskii2015} argue that mass transfer from giant donors may be much more stable than typically thought because a common envelope will not commence unless the donor exceeds its Roche lobe to such an extent that mass-loss occurs through the outer Lagrange points. They perform hydrodynamic modeling to determine the $q_\mathrm{crit}$ values relevant for this L2/L3 overflow scenario. They find that for evolved giants with deep convective envelopes, $q_\mathrm{crit}$ ranges from 1.5-2.2, twice the critical mass ratio predicted by polytrope models like \citet{Hjellming1987}. Based on these results, we run population synthesis models using $q_\mathrm{crit}=1.8$ for red giants and AGB stars, the approximate value found for giant stars in the mass range 1.0-2.0 \Msolar\ in these detailed simulations. This model is shown with the orange line in Figure~\ref{fig:MT_prescriptions}. 

\subsubsection{No Common Envelope}
\begin{figure*}
    \centering
    \includegraphics[width=.45\linewidth]{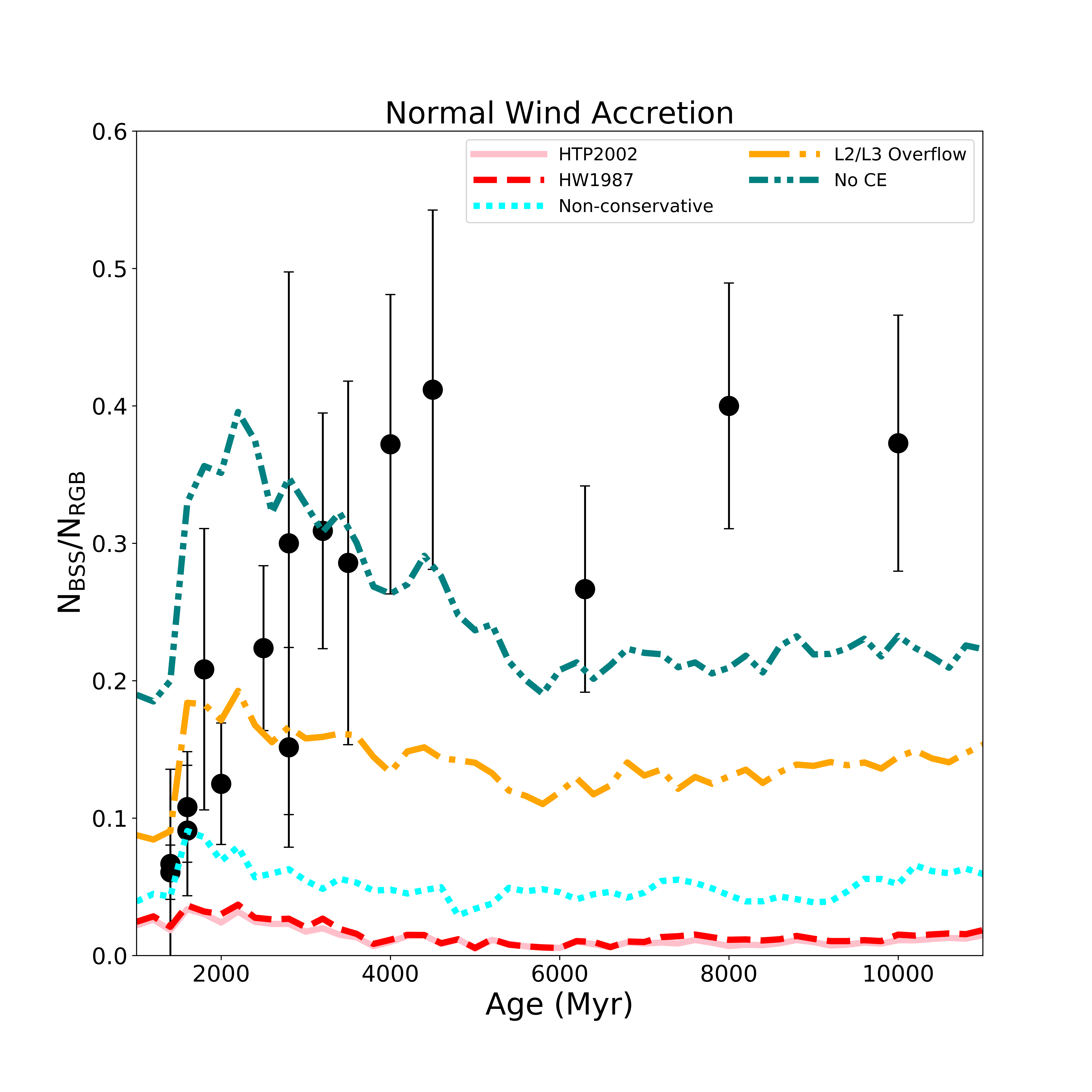}
    \includegraphics[width=.45\linewidth]{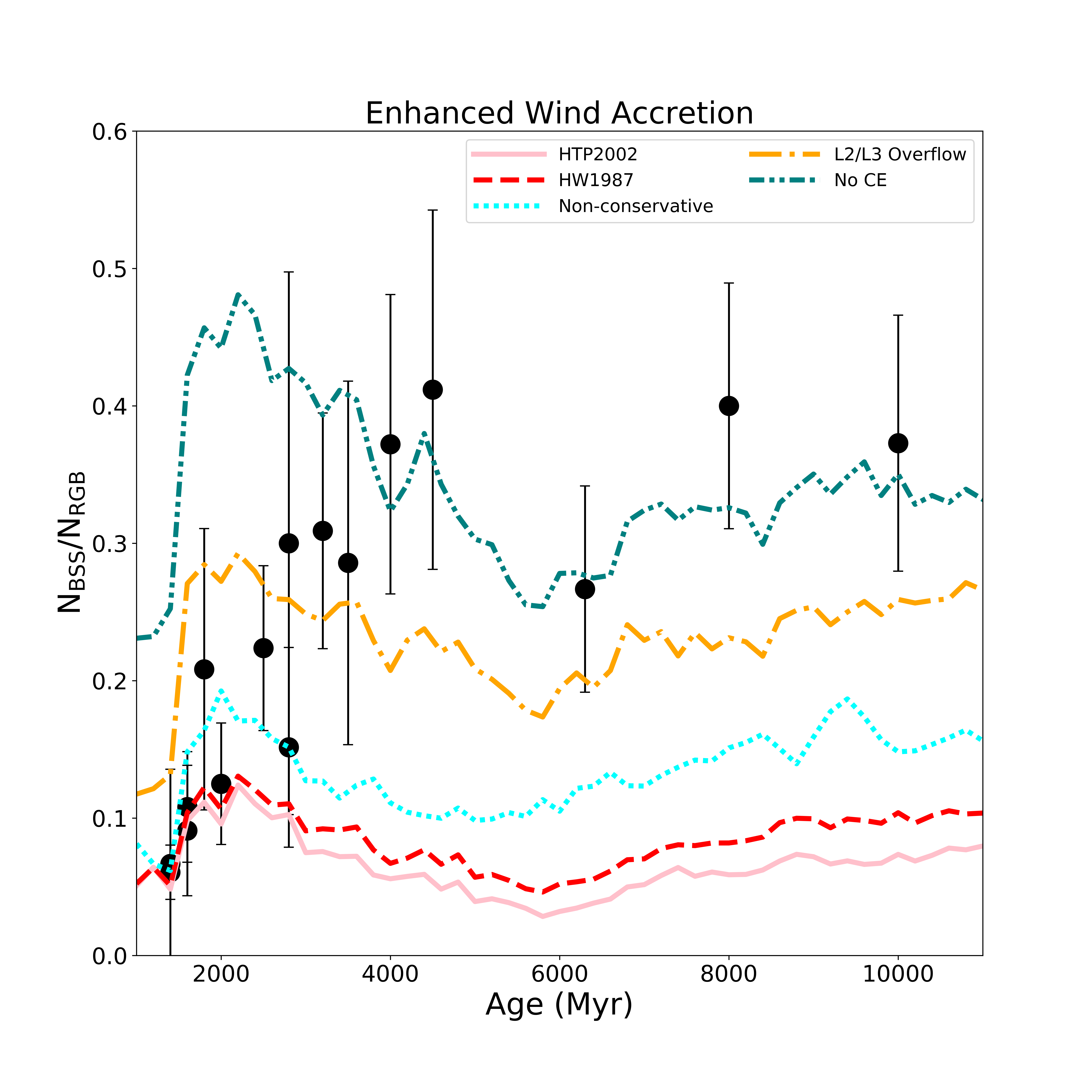}
    \caption{On the left, we show the observed ratio of the number of blue stragglers to the number of RGB stars in open clusters (black points) compared to~\texttt{COSMIC} population synthesis models. We show models with varying critical mass ratios for stability with colors as in Figure~\ref{fig:MT_prescriptions}. For these models, we use the default Bondi-Hoyle wind prescription. On the right, we show the same set of models, but increase the Bondi-Hoyle wind accretion by a factor of 10. We discuss this enhanced wind scenario in Section~\ref{section:wind}.}
    \label{fig:BSS_RGB_ratio}
\end{figure*}

We also consider a scenario in which mass transfer from an RGB or AGB donor always results in stable mass transfer. In this case, we set $q_\mathrm{crit}$= 100 for all red giant, red clump, AGB, and TP-AGB donor stars in~\texttt{COSMIC}. While this is not a realistic scenario, it sets an upper limit on how many blue stragglers may be created through mass transfer in~\texttt{COSMIC} if all binaries that interact during their giant evolution evolve through stable mass transfer. This model is not shown in Figure~\ref{fig:MT_prescriptions} (but the resulting blue straggler population is included in Figure~\ref{fig:BSS_RGB_ratio}). 

\subsection{Model Parameters}
\label{sectioN:modelparameters}
We implement each of the mass-transfer stability conditions discussed above in a~\texttt{COSMIC} population synthesis model. For all of our calculations, we implement ~\texttt{COSMIC}'s ~\texttt{multidim} sampling feature to select an initial population of binaries. We seed our models with an initial population of 100,000 binaries, all solar metallicity, and use the ~\texttt{delta\_burst} star formation model in which all stars are born at the same time. Binary mass ratios and orbital parameters are selected from the distributions reported in \citet{Moe2017}. This binary population is based on observations of large populations of binaries in the field, but investigations of open cluster binary populations have found similar results \citep{Geller2021, Geller2013, Geller2012, Duchene2013}). Very wide binaries that exist in the field will be dynamically disrupted in a cluster environment, but as these binaries are much wider than would be expected to interact and transfer mass, this difference should not impact blue straggler production.  

For all models, we lower the Reimer's wind mass loss coefficient from the default value of $\eta= 0.5$ to $\eta= 0.1$, consistent with asteroseismic measurements of RGB mass loss in older open clusters \citep{Miglio2012}. Wind mass transfer occurs via standard Bondi-Hoyle wind accretion in our models using the default Bondi-Hoyle accretion factor of $a_w= 1.5$. For all other parameters, we use the default ~\texttt{COSMIC} values. 


We note that this model population of 100,000 binaries is far larger than a typical open cluster in our sample, but by scaling the number of blue stragglers to the number of RGB stars in our models, the ratio $\frac{N_\mathrm{BSS}}{N_\mathrm{RGB}}$ is directly comparable to the observations. We discuss below how we define the blue straggler and RGB populations in our models.

\subsection{Model Blue Straggler Selection}
From our simulations, we select stars as blue stragglers that have temperatures $>5\%$ larger than the main sequence turnoff, and no more than a factor of 10 in luminosity below the main sequence turnoff. We use this luminosity cut to remove stars such as MS+hot WD binaries, and to be consistent with photometric cuts we make in our observational sample, and find that it excludes only a few possible blue stragglers in the models. 

We also count yellow stragglers in our model, selecting them to be at least twice the luminosity of the main sequence turnoff, with $T_\mathrm{eff} < 5\%$ larger than the main-sequence turnoff and $T_\mathrm{eff} > $ the maximum temperature found among subgiants in our model (~\texttt{kstar}=2) that have not been through any prior interactions. 

\subsection{Model RGB Selection}
Since our simulations contain only binaries, not single stars, and includes very wide binaries that would be past the hard-soft boundary in an open cluster, we cannot simply scale our BSS count to the number of RGB stars in our simulations as we do for the observations. Instead, we select all binaries from our simulation containing an RGB or red clump star (~\texttt{kstar} = 3 or 4) with orbital periods less than $10,000$ days. The spectroscopic binary fraction in both the field and in open clusters within this period domain is found to be $\sim25\%$ (\citealt{Geller2009, Milliman2014, Leiner2015, Moe2017}).  We multiply our selection of red giant binaries by 4 to calculate the total number of both RGB single and binary stars that would be in this population. We use this binary-fraction-corrected total to scale the blue straggler counts and compare to observations. 

\section{Results}\label{sec:results}

\subsection{Blue Straggler Fractions Using Different Mass Transfer Stability Prescriptions }
In the left panel of Figure~\ref{fig:BSS_RGB_ratio} we show the ratio of the number of blue straggler to the number of RGB stars in our 5 models compared to the observations.  All models under-produce blue stragglers in the older clusters. All models also show a similar shape, with a large increase in the ratio of blue stragglers to RGB stars between 1 Gyr and 2 Gyr, followed by a more gradual decrease and flattening. This shape is set by a few things: the formation frequency of blue stragglers, the lifetime of blue stragglers, and number of RGB stars at a given age. In the models, the steep rise from 1-2 Gyr is set primarily by a decreasing number of RGB stars, as at young ages RGB stars have longer lifetimes. After 2 Gyr in the models, the number of RGB stars flattens and the shape of this curve is dominated by the changing number of blue stragglers. The number of blue stragglers in the models declines slowly with increasing age. The lifetimes of blue stragglers in the older models is generally longer, since the blue stragglers are lower mass and have longer main-sequence lifetimes. Therefore, this decline reflects a decreasing production rate of blue stragglers. This decreasing production rate is because 1) more massive stars reach larger radii on the AGB, and therefore wider binaries evolve through mass transfer. 2) Wind accretion contributes more to younger models, and 3) mass transfer on the subgiant branch or Hertzprung gap is quite stable in BSE (\qcrit$=4.0$), and more blue stragglers form from this type of mass transfer in younger clusters. We discuss the results of each \qcrit\ prescription more specifically below. 

We find that the two most commonly used mass transfer prescriptions of \citet{Hjellming1987} and \citet{Hurley2002} (red and pink lines in Figure~\ref{fig:BSS_RGB_ratio}) produce very few, if any, blue stragglers in our simulations.  This is consistent with the youngest clusters in our study ($< 2$ Gyr), but not nearly as many observed in our older clusters. 

Adopting a non-conservative version of the \citet{Hjellming1987} such as discussed in \citet{Woods2011} does somewhat better. It produces approximately correct numbers of blue stragglers at ages $< 2$ Gyr, but still significantly under-predicts the number of cluster blue stragglers at ages $> 2$ Gyr, producing ratios $\sim0.05$ compared to the observationed ratio of 0.3-0.4. 
 
Our model using a constant $q_\mathrm{crit}=1.8$, approximately the value found in the L2/L3 overflow models of \citet{Pavlovskii2015}, produces a higher BSS to RGB ratio of $\sim0.15$ in clusters older than 2 Gyr. While much improved, this is only half the rate observed in older clusters. In clusters younger than 2 Gyr, this model produces approximately the observed number of blue stragglers. 

Finally, our model with \qcrit$= 100$, effectively sending all mass transferring binaries with RGB or AGB donors through stable mass transfer, produces a blue straggler to RGB ratio of $\approx25\%$ in the oldest clusters. This is still slightly below the observed fraction for clusters with ages $> 4$ Gyr. However, the model also overproduces blue stragglers in the youngest clusters ($< 2$ Gyr). 

It is interesting to note that even this model that eliminates common envelope does not produce the observed fraction of blue stragglers in old clusters, and we return to this point in Section~\ref{section:discussion}. 

In all models, we see that the shape of the distribution steeply rises to a peak at about 2 Gyr, before gradually flattening or declining at older ages. This is clearly in contrast to the observations, which steeply rise from 1-4 Gyr, before flattening.


 \subsection{HR Diagram Distribution of Blue Stragglers} 

\begin{figure*}
    \centering
    \includegraphics[width= .95\linewidth]{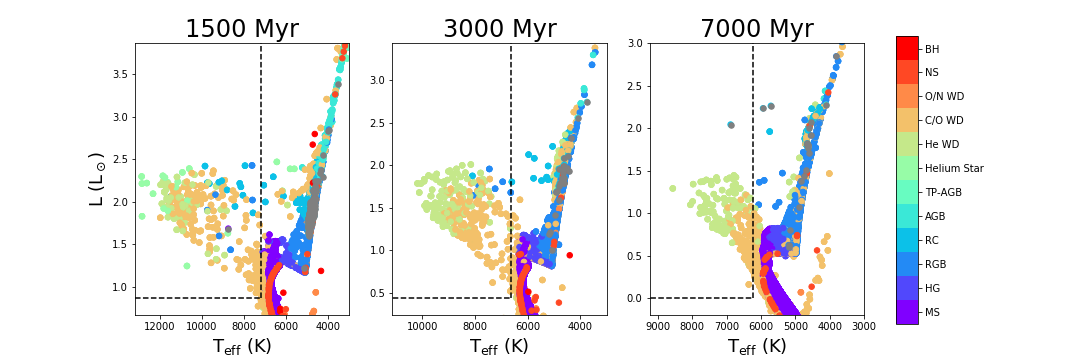}
    \includegraphics[width= .95\linewidth]{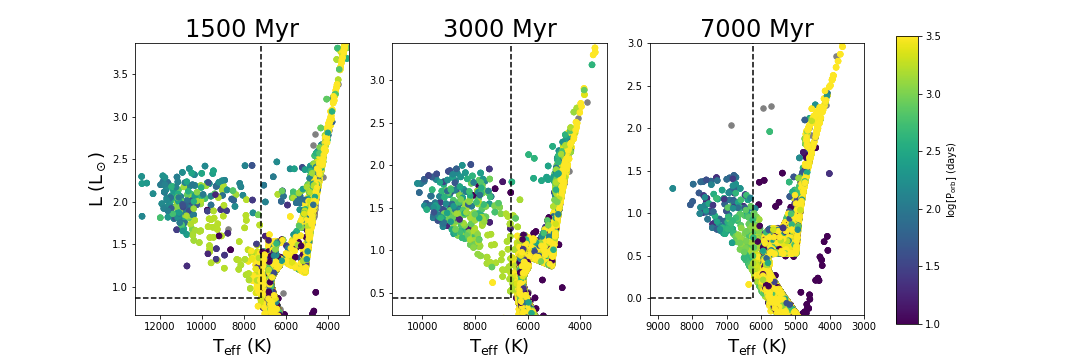}
    \caption{HR diagrams showing the results of COSMIC population synthesis simulations of 100,000 binaries using the L2/L3 overflow mass transfer stability prescription (\qcrit= 1.8). We show three snapshots in age: 1500 Myr (left), 3000 Myr (middle), and 7000 Myr (right). In the top panels we plot stars colored by the stellar type of the primary (i.e., the originally more massive star in our simulation). In the bottom panels we show the same simulations, but we color the stars by orbital period. Gray points indicate systems that are now single stars (i.e., due to a binary merger) and thus do not have orbital periods. The dashed lines indicate the boundaries of the blue straggler domain. We show the same plots for the other \qcrit\ prescriptions in the Appendix.}
    \label{fig:L2HR}
\end{figure*}

We produce some synthetic Hertzprung-Russell (HR) diagrams from these ~\texttt{COSMIC} runs for comparison to real clusters. Figure~\ref{fig:L2HR} shows an example HR diagram for the L2/L3 overflow model. For readability, the HR diagrams for the rest of the \qcrit\ prescriptions are given in the Appendix (Figure~\ref{fig:HTPcmd}-- Figure~\ref{fig:NoCEHR}). These plots show the results of COSMIC population synthesis runs for each of our stability criteria using 100,000 binaries at three different age snapshots: 1.5 Gyr, 3 Gyr, and 7 Gyr. In all plots we show two versions of these HR diagrams; one with points colored by the stellar type of the primary and one colored by the orbital period of the binary. We note that this sample size is much larger than the typical open cluster in our sample, which have typical stellar masses of a few thousand \Msolar. The total number of blue stragglers in these models is thus not directly comparable to a cluster, but provides a clearer picture of the distribution of blue straggler properties produced in each model.
 
 In nearly all cases, these HR diagrams show that the blue stragglers in our models have companions that are either Helium white dwarfs (green points) or carbon-oxygen white dwarfs (orange points). In the youngest models (1.5 Gyr plots), a few blue stragglers are still actively accreting and contain various types of giant primaries (blue or cyan points) or have been stripped of their hydrogen envelopes and are now Helium stars. 
 
 These HR diagrams also show that the \citet{Hurley2002} and \citet{Hjellming1987} prescriptions produce few blue stragglers, particularly at older ages.  Nearly all blue stragglers produced have low-mass Helium white dwarf companions in short-period orbits on of tens of days or less (see Figure \ref{fig:HTPcmd} and  \ref{fig:NWcmd}). This stands in contrast with the orbits determined for binary blue stragglers in old ($>1$ Gyr) open clusters, the majority of which have orbital periods of order 1000 days \citep{Geller2011, Milliman2014, Mathieu1986, Nine2020}. The sample of complete blue straggler orbits in most clusters is limited, and there are certainly cases of blue and yellow stragglers in these clusters with helium white dwarf companions and/or orbital periods of a few days (see, for example, \citealt{Gosnell2019, Landsman1997, Sandquist2003}), but they appear to be less common than the long-period blue stragglers.

 Our non-conservative prescription (Figure~\ref{fig:NCcmd}) produces more blue stragglers, and in particular produces more long-period blue stragglers with C/O white dwarf companions and orbital periods near 1000 days. This is particularly true in the 7 Gyr model snapshot where short-period systems are rare. Younger clusters still contain many shorter-period blue stragglers, and fewer long-period systems than at older ages. In general, the brightness of the blue stragglers is related to the orbital period of the system, with the brightest blue stragglers having He WD companions and shorter orbital periods of 10s or 100s of days, and the longer-period blue stragglers with C/O white dwarf companions appearing fainter and closer to the MSTO. Very short-period blue stragglers of just a few days are scattered throughout the blue straggler region. 
 
 The HR diagrams produced by the L2/L3 overflow condition (\qcrit= 1.8; Figure~\ref{fig:L2HR}) produce similar distributions to the non-conservative prescription, but produce many more blue stragglers. Again, these models produce blue stragglers with orbits ranging from a few to thousands of days, most with C/O white dwarfs or He white dwarf companions. A similar gradient in orbital period can be observed, with longer period blue stragglers appearing fainter than shorter-period blue stragglers. 
 
 The HR diagram produced by eliminating common envelope evolution (\qcrit= 100; Figure~\ref{fig:NoCEHR}) again produces lots of blue stragglers with orbits ranging from a few to thousands of days, most with Helium or C/O WD companions. The relationship between brightness and orbital period is less apparent, however, and blue stragglers near the MSTO have a mix of Helium white dwarf and C/O white dwarf companions and possess a range of orbital periods.

 

\subsection{Model Blue Straggler Mass Distributions}

\begin{figure*}
    \centering
    \includegraphics[width=.9\linewidth]{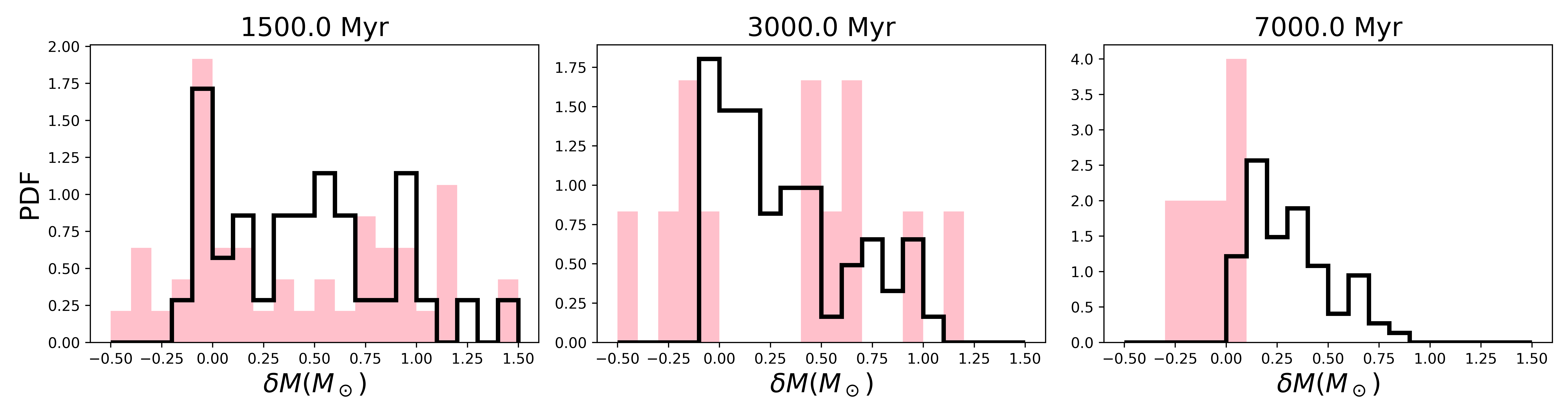}
    \vspace{.25cm}
    \includegraphics[width=.9\linewidth]{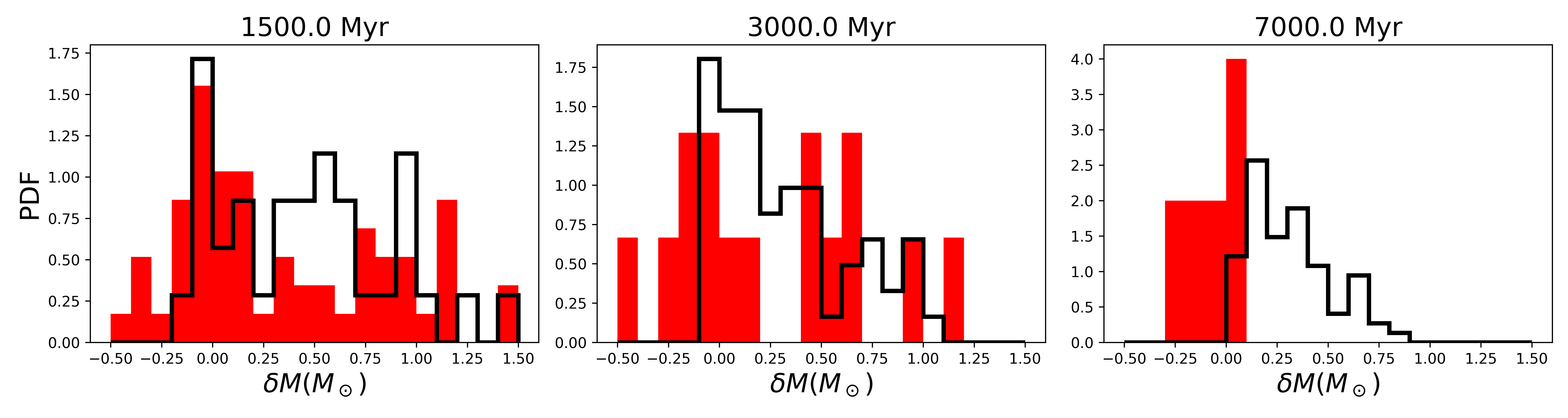}
    \includegraphics[width=.9\linewidth]{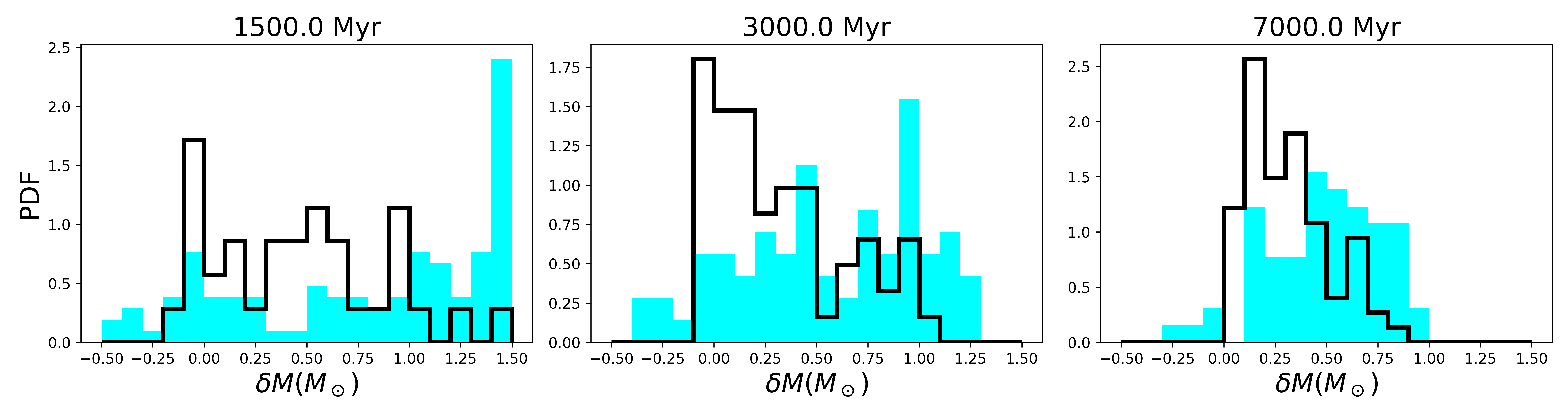}
    \includegraphics[width=.9\linewidth]{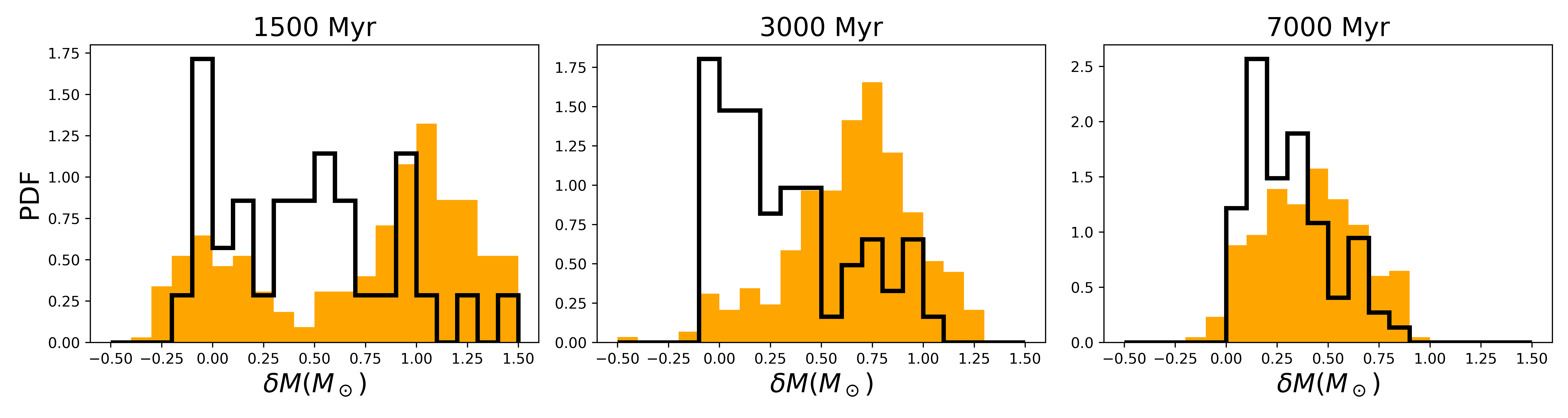}
    \includegraphics[width=.9\linewidth]{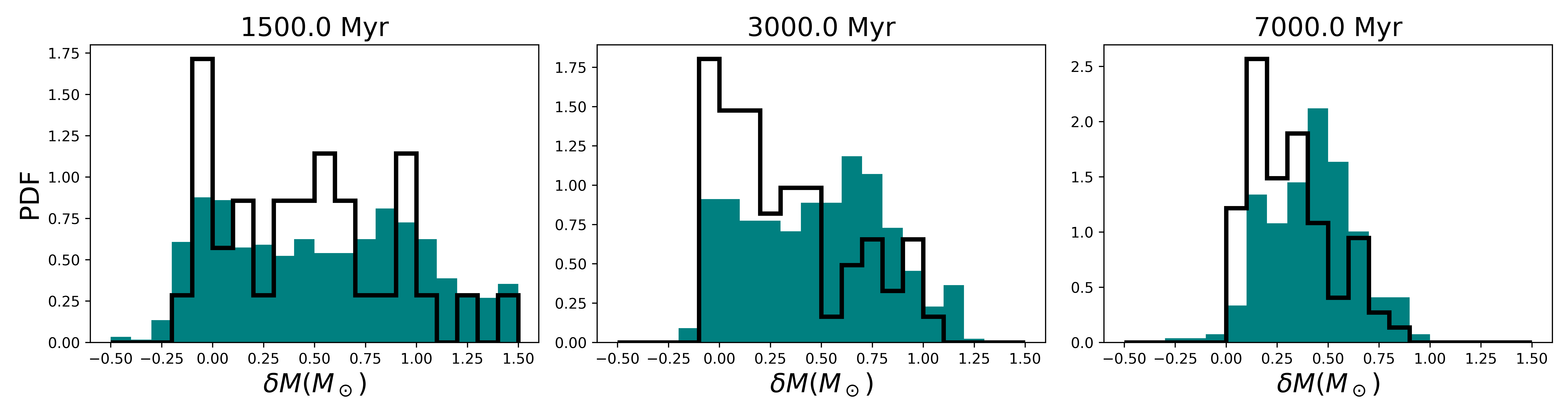}
    \caption{We show the mass distributions of blue stragglers produced in our models. On the x-axis, we plot the difference between the cluster turnoff mass and the blue straggler mass. On the y-axis we plot probability density distribution (PDF) of the blue straggler masses. Each row is a different model, colored as in Figure~\ref{fig:MT_prescriptions} (pink for \citet{Hurley2002}; red for \citet{Hjellming1987}; cyan for non-conservative; orange for \qcrit= 1.8; teal for \qcrit= 100). For comparison, we also plot the normalized number of blue stragglers from our Gaia cluster sample in black.}\label{fig:model_masshists}
\end{figure*}

In Figure~\ref{fig:model_masshists}, we show the distribution of masses in each of the snapshots discussed above: 1.5 Gyr, 3 Gyr, and 7 Gyr for each of our mass transfer prescriptions. 

We see that the \citet{Hurley2002} and \citet{Hjellming1987} produce similar mass distributions, with more blue stragglers produced near the MSTO and more massive blue stragglers produced more rarely. However, these models produce so few blue stragglers the comparison to the observed mass distributions does not yield much new insight. These models under-produce blue stragglers of every mass. 

The non-conservative prescription under-produces low-mass blue stragglers near the MSTO and over produces the more massive blue stragglers compared to the observed populations across all ages. This limitation is particularly noticeable in the youngest snapshot, where the models produce many massive blue stragglers more than 1.0 \Msolar\ above the turnoff. Blue stragglers this bright are rarely observed in the real clusters. Notably, the model mass distribution at 3.0 Gyr does show a paucity of blue stragglers near $\delta$M$= 0.5$, a feature also seen in the mass distribution of the observations. 

The \qcrit= 1.8 prescription produces a double-peaked mass distribution at 1.5 Gyr, which is not observed in the true 1-2 Gyr population. This model again over-produces very massive blue stragglers, and under-produces intermediate-mass blue stragglers with masses of a few tenths \Msolar\ above the turnoff. At 3 Gyr, the model under-produces blue stragglers near the MSTO and does not produce the double peaked mass distribution observed in the real clusters. At 7 Gyr, the mass distribution reproduces the observed distribution fairly well. If there is a small mass gap around 0.5 \Msolar (0.1 \Msolar\ in width) in the old clusters, it is not reproduced by the simulations. 

The no common-envelope prescription produces mass distributions very similar to the $q= 1.8$ prescription, under-producing low mass blue stragglers and overproducing higher mass blue stragglers.  

None of the models explored here perfectly reproduce the observed mass distribution of blue stragglers across age groups. The higher \qcrit\ models of $q= 1.8$ and $q= 100$ do the best job producing the observed mass distributions in old clusters, but both models produce too many high-mass blue stragglers and too few low-mass blue stragglers at young and intermediate ages. The non-conservative model is the only one which may produce a mass-gap like the one observed in the intermediate-age cluster, though the peak at lower BSS masses is smaller than observed. This model, too, underproduces low mass blue stragglers and overproduces high mass blue stragglers relative to the observations. 

Taken together, the HR diagrams and mass distributions of our models reveal two things about the BSS population models. 1) Default assumptions about mass-transfer stability far under-produce observed blue stragglers in old clusters. Even setting \qcrit=100 is unable to produce the observed numbers of blue stragglers in real old open clusters. 2) Models produce relatively more high mass blue stragglers than observed, but don't produce enough low-mass blue straggler near the turnoff. We discuss possible reasons for these discrepancies below.

 
\section{Discussion}
\label{section:discussion}
 \subsection{The Contribution of Stellar Dynamics, Collisions and Mergers}
 We have focused thus far on mass-transfer from RGB and AGB donors as the dominant blue straggler formation channel, but several other formation mechanisms have been proposed that likely contribute to the observed blue straggler population including stellar collisions, binary mergers, and interactions in triple star systems. 

\subsubsection{Stellar Dynamics}
Dynamics have long been thought to produce blue stragglers in dense cluster environments during stellar collisions \citep{Leonard1989}. $N$-body simulations of open clusters do produce blue stragglers via this channel \citep{Geller2013, Hurley2005}. A particularly relevant $N$-body model was done of NGC 188 \citep{Geller2013}, an old (6 Gyr) cluster in our sample with a large blue straggler population. This $N$-body simulation uses a \texttt{BSE} framework to perform stellar and binary evolution calculations, and thus uses much of the same input stellar physics as \texttt{COSMIC}. This model can therefore give us some insight into how the addition of dynamics might modify our results.

\citet{Geller2013} produce a mean of 6 blue stragglers in their cluster simulation, about 1/3 of their observed number of blue stragglers (20; \citealt{mathieu2015}) in NGC 188. At old ($> 3$ Gyr) ages, about 3 of these blue stragglers form via stellar collisions. At younger ages, fewer blue stragglers are produced through collisions. The simulations produce only $\sim 1$ collisional blue straggler at 1 Gyr, rising to $\sim 2$ blue stragglers at 2 Gyr, and finally $\sim 3$ blue stragglers by 3 Gyr. These results suggest dynamics could contribute $\lesssim$15\% (3/20) of the blue straggler population in our older clusters, and perhaps even fewer in the youngest clusters.

\subsubsection{Main Sequence Mergers}
Nearly all the blue stragglers in \citet{Geller2013} $N$-body models of NGC 188 not produced via collisions are produced via mergers of two main sequence stars. This amounts to 2-4 blue stragglers across all ages in their model.  Thus \citet{Geller2013} predict that mergers can account for $\sim$10-20\% of the blue stragglers in the clusters studied here, with the oldest clusters ($\geq 6$ Gyr) producing slightly fewer blue stragglers via mergers. 

\citet{Andronov2006} provide another analysis of the fraction of blue stragglers in open clusters formed from mergers, concluding that about 1/3 of the blue stragglers in open clusters result from main-sequence mergers. However, this estimate results from adopting an initial period distribution with a large fraction of short-period binaries, considerably larger than found in old clusters and the old field population. The number of mergers produced in a model is quite sensitive to assumptions about the number of these primordial close binaries. This choice likely explains the larger contribution from mergers than found in \citet{Geller2013}, who use an orbital period distribution more closely matching the observed orbital period distribution of NGC 188. 
 
We find that we rarely produce blue stragglers via a main-sequence merger channel in our COSMIC models, possibly because the initial period distribution we adopt from \citet{Moe2017} contains very few of the short-period binaries that would undergo mass transfer while on the main sequence (P$_\mathrm{orb} \leq 1.5$ days). Additional dynamics may be needed to shrink the orbits of some hard binaries for mergers to occur. We may also be able to produce a few more blue stragglers in our models via this merger channel by using a different magnetic braking law used (we use the law from \citealt{Ivanova2013} rather than the original \citealt{Hurley2002} prescription, which is the default for \texttt{COSMIC}), other prescriptions for angular momentum loss, or altering the stability criteria adopted for mass transfer from main-sequence or subgiant donors (for which we use the default \citealt{Hurley2002} settings).
 
We conclude that our models under-estimate the contribution of main-sequence mergers compared to some investigations in the literature. Considering the results of \citet{Geller2013} and \citet{Andronov2006}, mergers might make up an additional 10-30\% of the observed blue straggler populations. 
 
\subsubsection{Triple Systems} 
Triple systems, which are not included in our \texttt{COSMIC} models, may also contribute to the blue straggler populations. Kozai cycles plus tidal friction (``KCTF") in hierarchical triples may lead to the merger of the inner binary system, thus forming a blue straggler \citep{Ivanova2008, Perets2009}, and dynamically induced collisions may also be more likely in triple systems than in binaries (adding a wider tertiary star increases the cross section for interaction of the system). \citet{Geller2013} allow for the dynamical formation of triples in their NGC 188 simulation, and the model attempts to account for Kozai cycles and tidal friction. \citet{Geller2013} did not seed their canonical NGC 188 model with primordial triples, and find that insufficient triples form dynamically, as compared to observations.  In this canonical model, essentially no blue stragglers are formed from triples.  \citet{Geller2013} also produce a few simulations with varying numbers of primordial triples, whose masses and orbital periods were specifically chosen to facilitate blue straggler production at the age of NGC 188 and with the appropriate period distribution.  Even with the maximum number of possible triples (assigning every available binary in the appropriate mass range to a triple), \citet{Geller2013} find only 1 - 2 additional blue stragglers created above the canonical model. However, they also note that the efficiency of the KCTF mechanism in the \texttt{nbody6} code may be underestimated, and/or the implementation may need improvements.

Most short-period binaries (those that would lead to main sequence -- main sequence mergers) are found to have hierarchical triple companions, with the frequency of tertiary companions dropping off for inner binaries with wider orbits \citep{Raghavan2010, Moe2017}. Therefore, there is considerable overlap between those systems that may merge due to Kozai oscillations in triples, and those that may merge via a standard main-sequence mergers in close binaries as discussed above. The \citet{Perets2009} KCTF mechanism, therefore, does not boost blue straggler production substantially above the estimates of \citet{Andronov2006}. The important difference is that when these inner binaries merge, they will have a wide binary companion (most likely a low-mass main sequence star). 

We conclude that triples may enhance the numbers of stars produced through mergers or dynamics. However, this effect is already largely included in the estimates of the contribution from mergers and dynamics considered above. We do not find much evidence that including triples would significantly increase blue straggler production above these estimates, although the characteristics and dynamics of the triple population in clusters remains an open topic of research. 

\subsubsection{Correcting Straggler Production for Stellar Dynamics, Collisions and Mergers }
In the analysis above, we estimate that $\sim15$\% of blue stragglers may be produced by stellar collisions, and another $\sim10-30\%$ may arise from mergers. Overall, this means 25-45\% of all blue stragglers in our open cluster sample may arise from mergers and dynamics. This leaves 55-75\% of the blue straggler population to have formed via mass transfer from a giant donor. 

As we explain in the introduction, significant observational evidence points to mass-transfer as an important formation mechanism for blue stragglers in clusters and in the field. To revisit one important finding, \citet{Gosnell2014, Gosnell2015} use Hubble Space Telescope UV photometry to search for hot white dwarf companions to the blue stragglers in NGC 188. They detect 7 white dwarf companions, all hot enough to have formed in the last 400 Myr. Given that older, cooler white dwarfs are undetectable with their method, they estimate that with an incompleteness correction, two-thirds of the NGC 188 blue stragglers have white dwarf companions. This implies that 2/3 of the blue straggler population of this cluster formed via mass transfer from a giant donor. This study only looks at the population of one open cluster, but it is the most direct measurement available for the fraction of blue stragglers formed through mass transfer. This result is in line with our estimate from theoretical models that 55-75\% of blue stragglers formed from mass transfer.

If we assume that the blue stragglers formed in our COSMIC models account for only 55\% of the real blue straggler population, then the L2/L3 overflow model and the no CE model are consistent with the observed blue straggler fraction found in old clusters, but significantly over-predict the number of blue stragglers found in 1-2 Gyr clusters (Figure~\ref{fig:BSS_RGB_ratio}). The non-conservative model would still underestimate the number of blue stragglers in old clusters by a factor of 3, but would be approximately correct for the 1-2 Gyr clusters. The \citet{Hurley2002} and \citet{Hjellming1987} prescriptions still produce far too few blue stragglers at all ages. 

The underproduction of blue stragglers through mass transfer was also noted by \citet{Geller2013} in their NGC 188 model, that used the \citet{Hjellming1987} \qcrit\ equation.  \citet{Geller2013} also investigate the effects of using various \qcrit\ approaches, though not directly in their canonical NGC 188 $N$-body model, to potentially explain both the model's lack of blue stragglers and also the model's over-abundance of long-period ($\sim$100-1000 day) circular main-sequence -- white dwarf post-CE binaries (which are not observed in true binary populations).  \citet{Geller2013} discuss that if these spurious post-CE binaries instead were meant to become blue stragglers, the NGC 188 $N$-body model would produce roughly the correct number and orbital period distribution of the observed NGC 188 blue stragglers.  Our no CE model investigates this possibility as well, and indeed does produce the predicted $\sim$50\% of blue stragglers through mass transfer for old open clusters.

We conclude that correcting for other production mechanisms may bring the L2/L3 model and no CE model into agreement with the total number of blue stragglers produced in older clusters, but using the same correction in young clusters results in too many blue stragglers. Perhaps the relative number of blue stragglers produced via collisions and mergers varies with age.  (Indeed the \citealt{Geller2013} NGC 188 model suggests that collisions may only be relevant for clusters at $\gtrsim$2-3 Gyr, and that the merger rate may drop off toward older clusters, though these two effects nearly balance out over time.)  

We note that the 5 stability criteria investigated here are not exhaustive (e.g. \citealt{Chen2008, Belczynski2008, Ge2020}, and others), and other models may yield results that better match the observed number of blue stragglers. However, there are also other parameters in these models that may need adjustment. We discuss some possible important parameters below.

\subsection{Additional Mechanisms that can Contribute to the Number of Blue Straggler}

 \subsubsection{Wind Mass Transfer} 
 \label{section:wind}


 \begin{figure*}
    \centering
 \includegraphics[width=.9\linewidth]{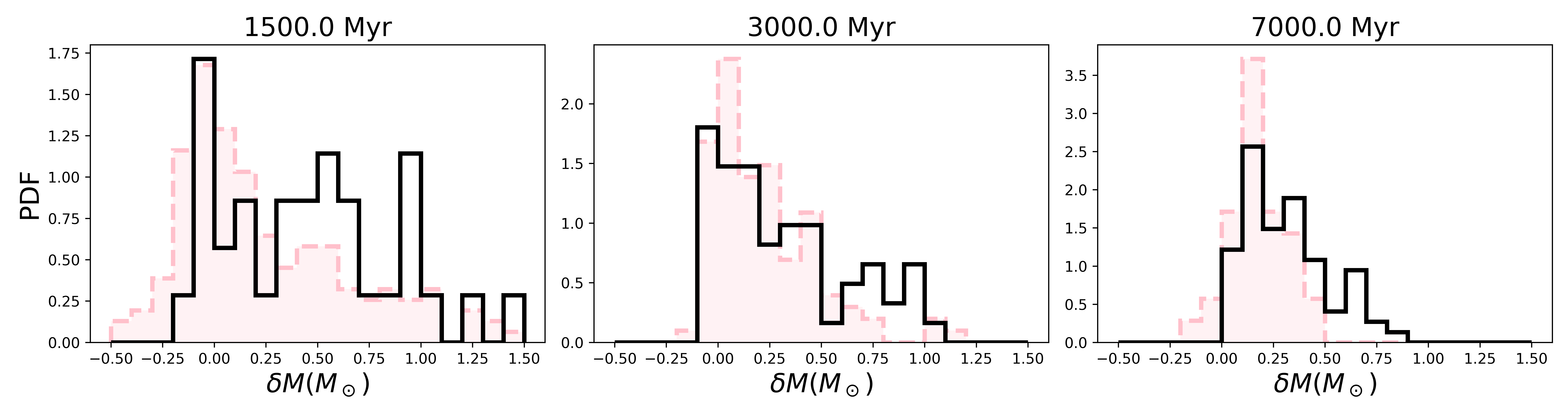}   
  \includegraphics[width=.9\linewidth]{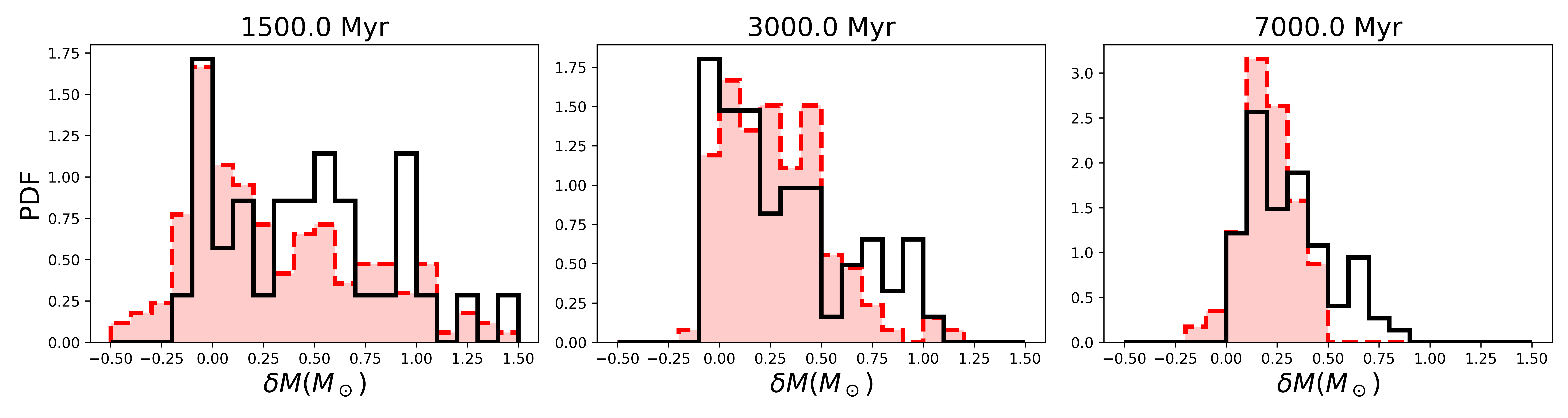}   
    \includegraphics[width=.9\linewidth]{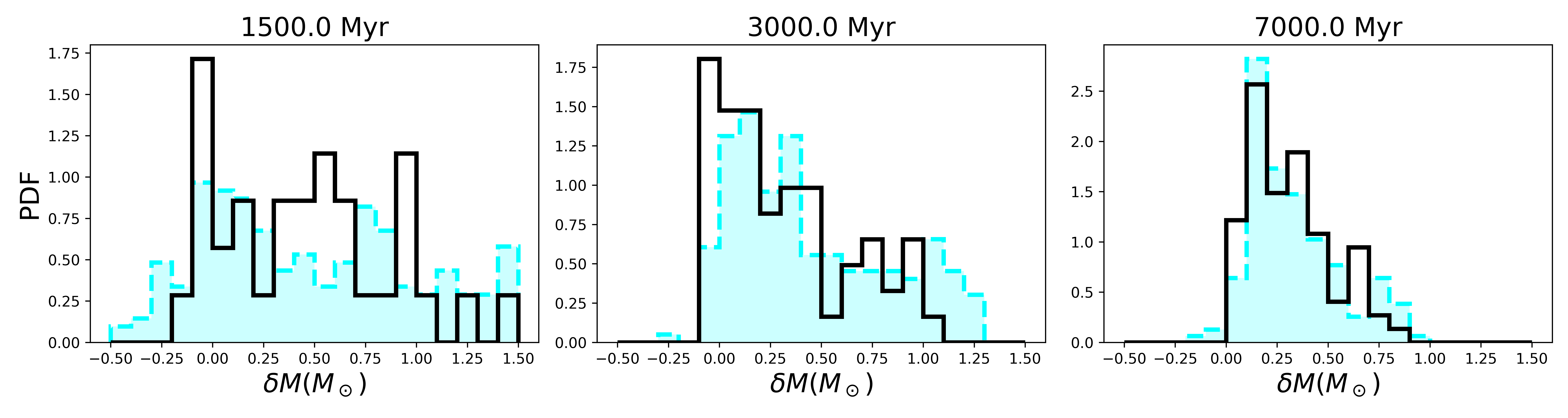}    
    \includegraphics[width=.9\linewidth]{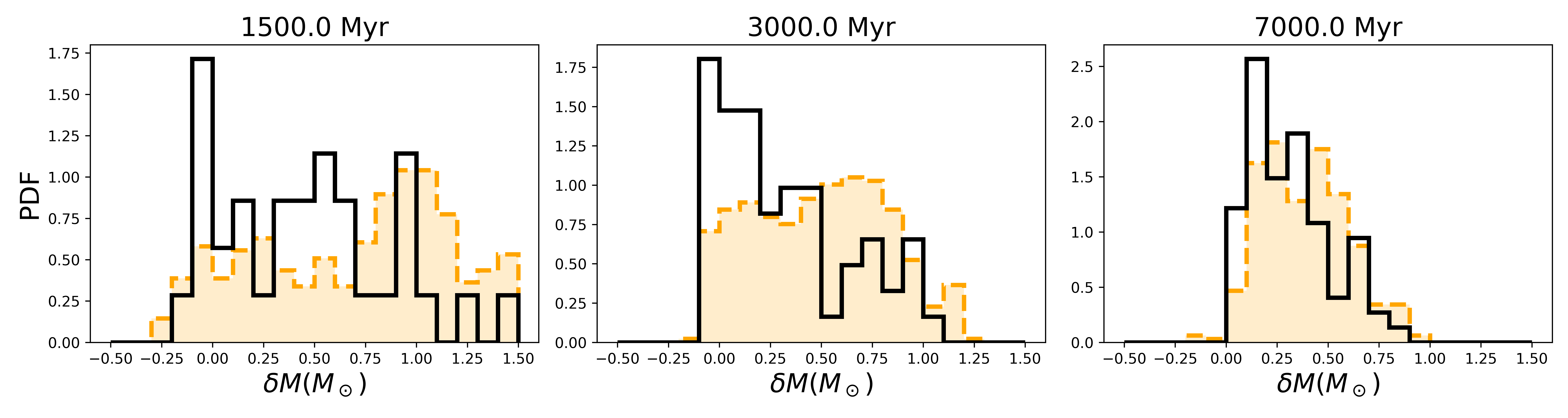}
     \includegraphics[width=.9\linewidth]{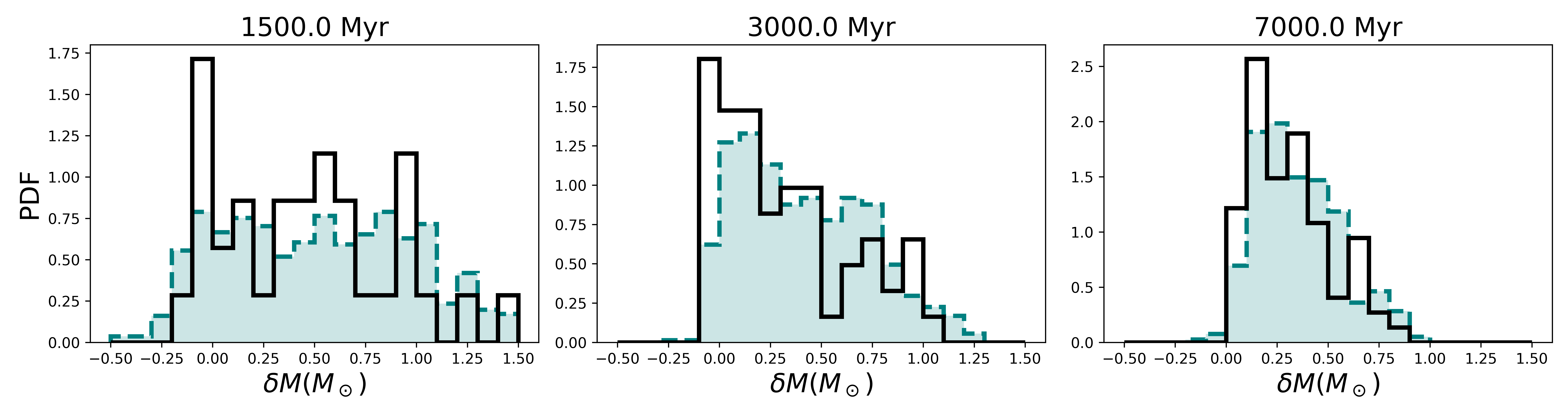}   
    \caption{Same as Figure~\ref{fig:model_masshists}, but showing model blue straggler mass distributions when increasing the Bondi-Hoyle wind accretion by a factor of 10.}
    \label{fig:model_masshists_wind}
\end{figure*}
Recent modeling has showed that mass transfer may occur at wider separations than expected through typical Roche lobe overflow \citep[e.g.][]{Abate2013, Chen2017, Mohamed2007}. This mechanism has been suggested to be particularly important in creating barium stars and carbon enhanced metal poor stars with s-process enrichment (CEMP-s).  These stars are relatives of blue stragglers and are identified as products of mass-transfer because of observed enhancements in s-process elements like barium that are likely the result of accretion of s-processed enriched material from an AGB donor star \citep{Boffin1988}. 
 
 In the context of s-process enhanced stars like these, hydrodynamic models show that the slow wind driven from the star during the final phases of AGB evolution can, in some circumstances, be accreted by the companion with efficiencies above typical Bondi-Hoyle accretion \citep[e.g.][]{Abate2013, Saladino2019}, in some cases reaching efficiencies as high as 50\% \citep{Mohamed2007}. This is often termed ``Wind  Roche  Lobe overflow". 
 
 \texttt{COSMIC/BSE} does not include a Wind Roche Lobe overflow mechanism. Our models use only standard Bondi-Hoyle accretion, and thus may underestimate the number of wide blue straggler binaries forming as a result of wind accretion from an AGB companion. 
 
To test the impact of our wind assumption on the resulting blue straggler populations, we re-ran all our models using an enhanced Bondi Hoyle wind accretion in which we raise the constant by a factor of 10 ($a_w= 15$, raised from $a_w= 1.5$). This value is unrealistically high, but we use this case to test the impact additional wind accretion could have on the blue straggler population. All other parameters of the Bondi-Hoyle accretion are kept at the BSE defaults, which are described in \citet{Hurley2002}. We show plots of the resulting mass distributions and BSS to RGB ratios Figures ~\ref{fig:model_masshists_wind} and the right panel of Figure ~\ref{fig:BSS_RGB_ratio}, respectively.)

 We find the more efficient wind accretion increases the BSS to RGB ratio by $\sim5-10$\%. Increasing wind accretion also leads to the production of more low-mass blue stragglers, bringing the theoretical blue straggler mass distributions closer to the observed distributions, though models using high \qcrit\ still overproduce high-mass blue stragglers (Figure~\ref{fig:model_masshists_wind}). The non-conservative prescription with wind matches the observed mass distribution particularly well, showing a double peaked distribution in intermediate and old populations like the observations (see Figure~\ref{fig:model_masshists_wind}, cyan plots). 
 
We suggest that wind mass transfer may be a viable mechanism to produce more blue stragglers close to the turnoff. However, in our \texttt{COSMIC} models wind mass transfer seems to increase the population of blue stragglers as much or more in the young clusters as in the old clusters. A larger increase in blue straggler production in the old clusters is needed to resolve the discrepancy with observations. A more careful treatment of wind mass transfer in population synthesis models is needed to examine this possibility more closely.  



\subsubsection{Binary ``Twins" and Blue Straggler Lifetimes}
  
 \begin{figure*}[hbt]
     \centering
     \includegraphics[width= .9\linewidth]{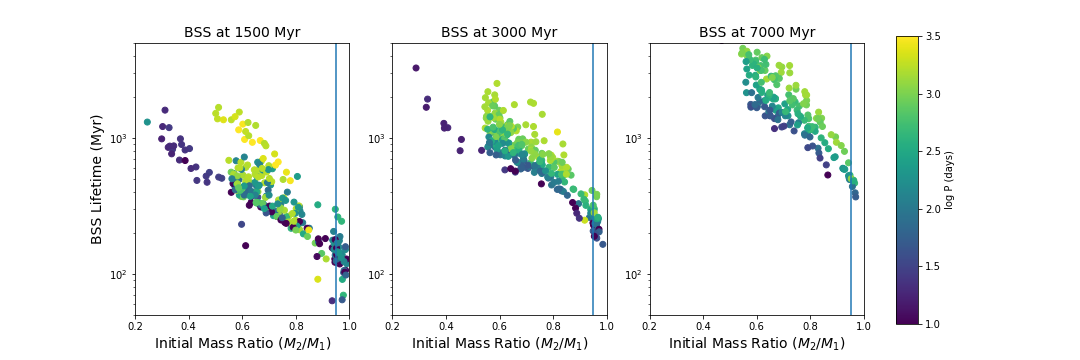}
     \caption{We show the blue straggler lifetime as a function of the initial binary mass ratio for three different time snapshots in our L2/L3 overflow model (\qcrit= 1.8 model): 1500 Myr (left), 3000 Myr (middle) and 7000 Myr (right). The colors indicate the initial orbital period of the binary that formed the blue straggler. These give an indication of the type of Roche lobe overflow that formed the blue straggler: $\sim1-10$ day periods binaries (purple) go through Roche lobe overflow beginning on the main sequence or subgiant branch, periods of a few hundred days (blue) begin RLO on the RGB, periods of $\sim1000$ days (green) go through RLO on the AGB, and periods larger than $\sim1000$ (yellow) days form via Bondi-Hoyle wind accretion.  We mark a mass ratio of 0.95 with the blue line. Blue stragglers that formed from twin binaries fall to the right of this line.}
     \label{fig:lifetimes}
 \end{figure*}

 In our $1-10$ Gyr models, the binary systems that will interact and transfer mass will have primaries with masses of $1-2$ \Msolar\ and orbital periods $\lesssim$ 1000 days. In this mass and period domain,  \citet{Moe2017} find that there is a substantial excess of binary systems with mass ratios close to unity ($0.95 < \frac{M_2}{M_1} < 1.0$). Because of the excess of these binary ``twins", and because a mass ratio near 1 is predicted to be stable over a large range of our models (Figure ~\ref{fig:MT_prescriptions}), these systems represent a considerable fraction of possible blue straggler progenitors. For example, among initial binaries in our simulations with $P_\mathrm{orb} < 1000$ days and $1.0< M_1< 2.0$, 20\% of these systems are twins. In the L2/L3 overflow model (\qcrit=1.8), twins represent 42\% of the initial binaries in our sample that are likely to go through stable Roche lobe overflow because they have mass ratios $\frac{M_1}{M_2} < 1.8$, primaries with $1< M < 2$ \Msolar\ and P$_\text{orb}< 1000$ days. In the non-conservative model, this fraction is even higher. 
 
In our models we find, however, that twins often never form blue stragglers, particularly at old ages. At 7 Gyr, only 1/3 of the twin population evolves through a blue straggler phase, and the blue straggler lifetime is only a few hundred Myr. For the majority of mass-transferring twins at 7 Gyr, the secondary begins to evolve off the main sequence before or during Roche lobe overflow, usually forming a single overmassive giant via a merger. As an example, we show the blue straggler lifetime as a function of the initial mass ratio of the binary in Figure~\ref{fig:lifetimes} for three different snapshots in our L2/L3 overflow model. Note the paucity of blue stragglers formed from binary twins, particularly in the oldest snapshot, despite the large progenitor population.
 
 These results indicate that we are quite sensitive to the initial parameters of twin binaries. For example, we assume that binary mass ratios of twins are distributed uniformly between 0.95 and 1.0. If the overabundance of twins began at $\frac{M_2}{M_1}= 0.9$ instead of $0.95$, we would produce more blue stragglers from twin binaries in old clusters. Similarly, if the population favors mass ratios closer to 0.95 rather than being uniformly distributed, we would produce more blue stragglers. These granular details of the twin population are beyond the precision of current studies, but relevant to the formation of blue straggler stars.
 
 Furthermore, \texttt{BSE} simply assumes that the fractional lifetime remaining for the accretor star will be preserved once it gains mass and becomes a blue straggler. If the blue straggler is rejuvenated at formation because more hydrogen is mixed into the core, the lifetime of the blue stragglers could be significantly longer. This rejuvenation is particularly important to understand in blue stragglers forming from twin binaries, where the remaining fractional lifetime of the blue straggler is small without including some rejuvenation (Figure~\ref{fig:lifetimes}). The amount of rejuvenation in blue stragglers is uncertain, and depends on the initial and final mass of the accretor, the rotational evolution of the system, and the amount of internal mixing, but it could be substantial. For example, \citet{Sun2021} create a detailed stellar evolution model of a blue straggler in NGC 188 (6 Gyr). The system has an initial mass ratio of $q= 0.85$, goes through non-conservative mass transfer, and experiences modest core hydrogen rejuvenation. The resulting blue straggler has a remaining main-sequence lifetime of $\sim 2$ Gyr. The lifetime of this blue straggler in our models would be $\lesssim1$ Gyr (Figure~\ref{fig:lifetimes}, right panel). A model for this blue straggler that included rapid rotation and a more detailed consideration of the internal mixing could experience even more rejuvenation.
 
 The twin fraction from \cite{Moe2017} is also higher than is found in some open cluster studies (e.g. \citealt{Geller2013}). If, for example, we removed the twin excess entirely and re-distributed these stars according to the power-law distribution in \citet{Moe2017}, we could produce more blue stragglers in most models. We expect this effect would be small, however, because the stars would be distributed over a wide range of initial mass ratios, most of which cannot form blue stragglers. This would raise the BSS to RGB ratio by a few percent in our \qcrit= 1.8 and \qcrit= 100 models, and less in other stability models.

\subsubsection{Mass Transfer Efficiency} 
 
 \begin{figure*}
     \centering
     \includegraphics[width=0.95\linewidth]{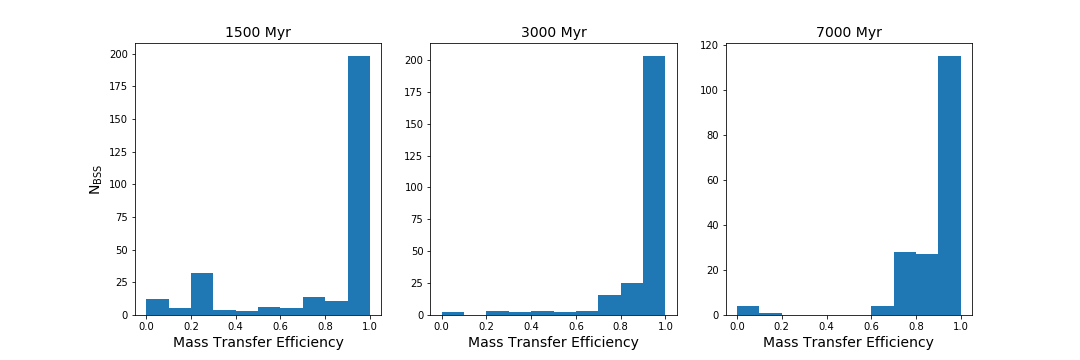}
     \caption{For our L2/L3 overflow model, we show the overall efficiency of the mass transfer (i.e. fraction of material lost by the donor that is accreted by the companion) that formed the blue stragglers. We calculate this simply by dividing the combined mass of the blue straggler binary by  the combined mass of the progenitor binary. This mass may have been lost during RLO or via a wind. We show these plots for blue stragglers in 3 different snapshots: 1.5 Gyr (left), 3 Gyr (middle) and 7 Gyr (right).}
     \label{fig:MTefficiency}
    \vspace{1cm}
 \end{figure*}

 As we state in Section~\ref{section:nonconservativeMT}, \texttt{COSMIC/BSE} generally assumes mass transfer will be conservative (i.e. all of the mass lost by the donor star during Roche lobe overflow is accreted by the companion). Nearly all the blue stragglers created in our models are produced via nearly conservative mass transfer, except in the case where the blue straggler is formed entirely via Bondi-Hoyle wind accretion (see Figure~\ref{fig:MTefficiency}). This is, of course, inconsistent with our non-conservative and L2/L3 overflow prescriptions, as these stability predictions assume significant material can be lost during the mass transfer process. Changing \qcrit\ can change the total number of blue stragglers produced by our models, but if nearly all these blue straggler form via conservative mass transfer, changing \qcrit\ does not change the mass distribution of the blue stragglers. Incorporating non-conservative mass transfer into population synthesis models, then, is likely required to produce realistic blue straggler mass distributions.

A straight-forward population synthesis prescription that self-consistently calculates \qcrit\ and mass transfer efficiency has not been developed. Nevertheless, non-conservative mass transfer is probably common when forming blue straggler stars. \citet{Gosnell2019} show that 2 post-mass-transfer blue stragglers in NGC 188 both likely formed via non-conservative mass transfer. Their formation history for one of the brightest blue stragglers in the cluster requires an overall mass transfer efficiency of just $60\%$, suggesting even the most massive blue stragglers in a cluster are not forming via fully conservative mass transfer. 

Allowing for lower mass transfer efficiencies during Roche lobe overflow could reduce the number of high mass (bright) blue stragglers and increase the number of lower-mass blue straggler (closer to the turnoff), bringing the mass distributions of the model blue stragglers in line with the observed distributions (Figure ~\ref{fig:masshist}). 

We also note that lower mass blue stragglers would also have longer lifetimes than higher mass blue stragglers, so this modification could also increase the number of blue stragglers stars in our models by increasing blue straggler lifetimes (Figure~\ref{fig:lifetimes}). This would be particularly relevant for the blue stragglers forming from twins we discuss in the previous section, since twin binaries form the most massive blue stragglers with the most depleted hydrogen cores. We suggest allowing that non-conservative mass transfer is an essential step in improving population synthesis models.

\section{Summary}\label{sec:summary}
We use Gaia DR2 to determine memberships for a sample of 16 nearby open clusters with ages ranging from 1.4-10 Gyr. This age range corresponds to a range in main-sequence turnoff masses of 2.0-1.0 \Msolar. 

We compare the observed blue straggler populations between clusters, finding that there is a steep rise in the ratio of blue stragglers to RGB stars in our clusters up to an age of $\approx4$ Gyr, at which point the ratio flattens to $\frac{N_\mathrm{BSS}}{N_\mathrm{RGB}} \approx 0.35$. This result indicates that old clusters produce more blue stragglers than young clusters, and/or that their blue straggler populations are longer lived. We compare this observed relationship to ~\texttt{COSMIC} population synthesis models using five different prescriptions for the critical mass ratio for mass-transfer stability (\qcrit). We find that all options under-produce blue stragglers in old ($> 3$ Gyr) clusters, even under the assumption of a very large \qcrit=100.0, in effect assuming all mass-transfer is stable and avoids common envelope evolution. 

We discuss other factors that may contribute to the number of observed blue stragglers, including blue stragglers produced via stellar collisions or mergers in triple systems. These other production mechanism could account for 25-45\% of blue straggler productions in clusters, making up for some of the discrepancy between models and observations in number blue stragglers. 

We also find that in aggregate, the observed blue straggler populations in clusters older than 2 Gyr are split into two groups: a bright group of blue stragglers 0.6-1.0 \Msolar\ above the MSTO and a fainter group 0.0-0.4 \Msolar\ above the MSTO. In these clusters, there is an observed paucity of blue stragglers at $\sim0.5$ \Msolar above the turnoff.  The faint blue stragglers are the majority group, accounting for 70\% of the overall blue straggler population summed over all clusters. 

This mass distribution is at odds with the results of most of our population synthesis models, which produce more bright blue straggler than observed, regardless of stability condition, and show no mass gap. We find more promising agreement in blue straggler mass distributions when adding in enhanced Bondi-Hoyle accretion that produces fainter blue stragglers via highly non-conservative wind mass transfer, particularly when using our non-conservative stability prescription (i.e., the cyan histograms in Figure~\ref{fig:model_masshists_wind}). 

We therefore suggest one explanation for the two groups of blue stragglers we observe in old clusters  could be that stable, nearly conservative mass transfer forms the brighter group of blue stragglers, whereas non-conservative mass transfer (e.g., via an AGB wind) forms the fainter group of blue stragglers. Alternatively, perhaps the two groups form via different channels (e.g., the brighter group forms from RGB mass transfer, the fainter group from AGB mass transfer). More detailed studies of blue straggler orbital properties to compare with model predictions could offer more insight into this puzzle.

We conclude that current population synthesis models have difficulty producing the observed blue straggler production rates and mass distributions in open clusters. It is clear that the standard prescriptions of \citet{Hjellming1987} or \citet{Hurley2002} fail to produce reasonable numbers of blue stragglers and a higher \qcrit\ should be used in population synthesis.  Using a substantially higher \qcrit\ value such as we use for our L2/L3 overflow model produces more realistic blue straggler counts, but over-produces bright blue stragglers relative to near-turnoff blue stragglers. These \qcrit\ prescriptions are not exhaustive, and implementing new \qcrit\ prescriptions based on model grids of detailed stellar structure calculations may reveal prescriptions that perform better (e.g. \citealt{Ge2020}). 

However, our results indicate that \qcrit\ is not the only parameter that needs adjustment in population synthesis in order to match the observed blue straggler populations. We suggest in particular that the mass distributions of blue straggler stars indicate  non-conservative mass transfer is important to their formation, and allowing for non-conservative mass transfer in population synthesis is a necessary ingredient to produce more realistic blue straggler populations. Coupling one of the more stable \qcrit\ prescriptions (e.g. the L2/L3 model) with a self-consistent calculation of the mass transfer efficiency may yield results more consistent with observations. Formation of blue stragglers via a highly non-conservative mass-transfer channel such as wind accretion may also be important.

In addition, core hydrogen rejuvenation in blue stragglers would increase blue straggler lifetimes, and a more careful treatment of this effect may be needed to produce observed quantities of blue stragglers in open clusters. We leave a detailed investigation of these factors to future modeling work.

Blue stragglers are one window into the outcome of mass transfer processes in low mass binaries. They form from the first mass transfer event in a binary's evolution, leaving a white dwarf -- blue straggler binary. These systems may later go on to form many types of interesting astrophysical binaries-- e.g. double white dwarfs, low-mass white dwarf mergers and exotic transients, X-ray binaries, and others. We demonstrate in this paper the difficulty of reproducing the observed blue straggler populations using common implementations of mass-transfer physics in population synthesis codes, and show that the blue straggler populations can provide insights into the mass-transfer process. Improving these mass-transfer models is important not only in understanding blue straggler formation, but in understanding a wide variety of post-mass-transfer binaries.

\acknowledgements
EML is supported by an NSF Astronomy and Astrophysics Postdoctoral Fellowship under award AST-1801937.This work has made use of data from the European Space Agency (ESA) mission
{\it Gaia} (\url{https://www.cosmos.esa.int/gaia}), processed by the {\it Gaia} Data Processing and Analysis Consortium (DPAC,
\url{https://www.cosmos.esa.int/web/gaia/dpac/consortium}). Funding for the DPAC has been provided by national institutions, in particular the institutions participating in the {\it Gaia} Multilateral Agreement. The authors also thank the anonymous referee for helpful comments on this work.

\bibliographystyle{apj}
\bibliography{citations}

vspace{10cm}

\clearpage
\begin{appendix}

In this Appendix, we show CMDs for all the open clusters in our sample (Figure~\ref{fig:allclustercmds}). We also show HR diagrams for the COSMIC population synthesis models assuming $q_\text{crit}$ prescriptions of \citet{Hurley2002} (Figure~\ref{fig:HTPcmd}), \citet{Hjellming1987} (Figure~\ref{fig:NWcmd}), our non-conservative prescription (Figure~\ref{fig:NCcmd}) and $q_\text{crit}= 100$ (Figure~\ref{fig:NoCEHR}). For more details about these models and HR diagrams see see Sections 4.1 and 5.2 in the main text. 

\renewcommand{\thefigure}{A.\arabic{figure}}

\setcounter{figure}{0}

\vspace{1cm}

\begin{figure*}[hbt]
    \centering
    \subfigure[]{\includegraphics[width=.3\linewidth]{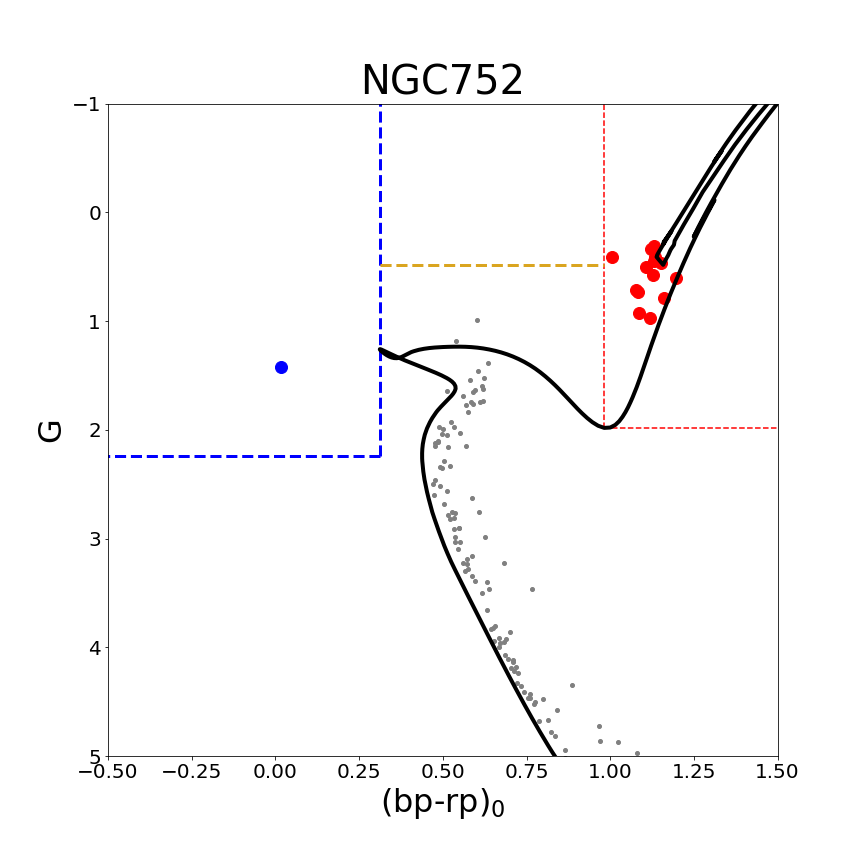}}
    \subfigure[]{\includegraphics[width= .3\linewidth]{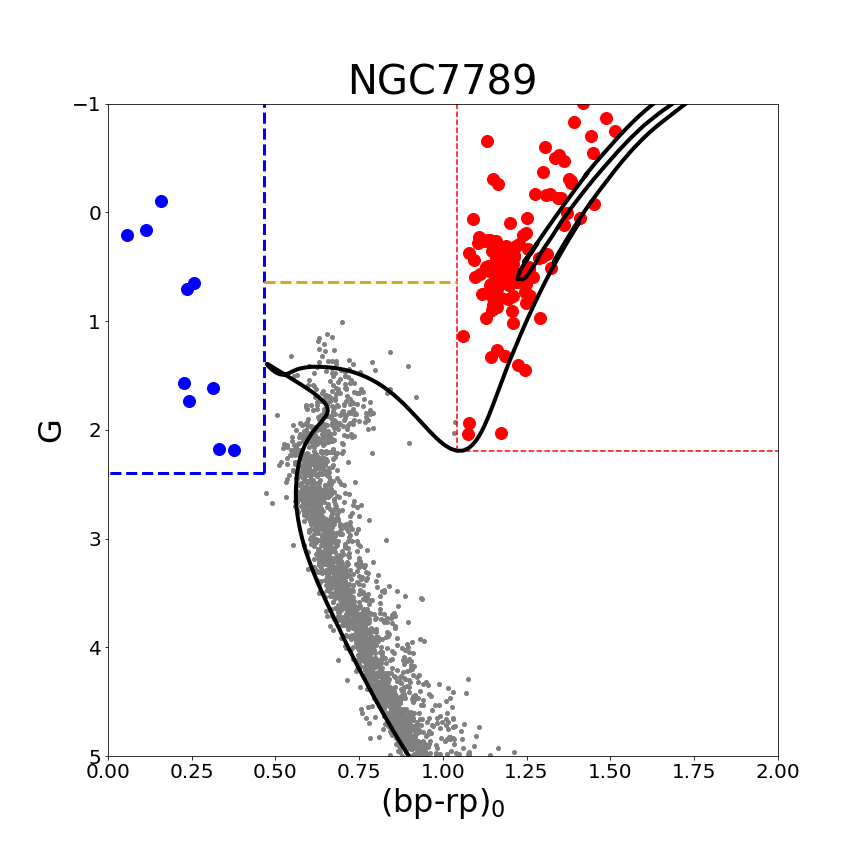}}
    \subfigure[]{\includegraphics[width=.3\linewidth]{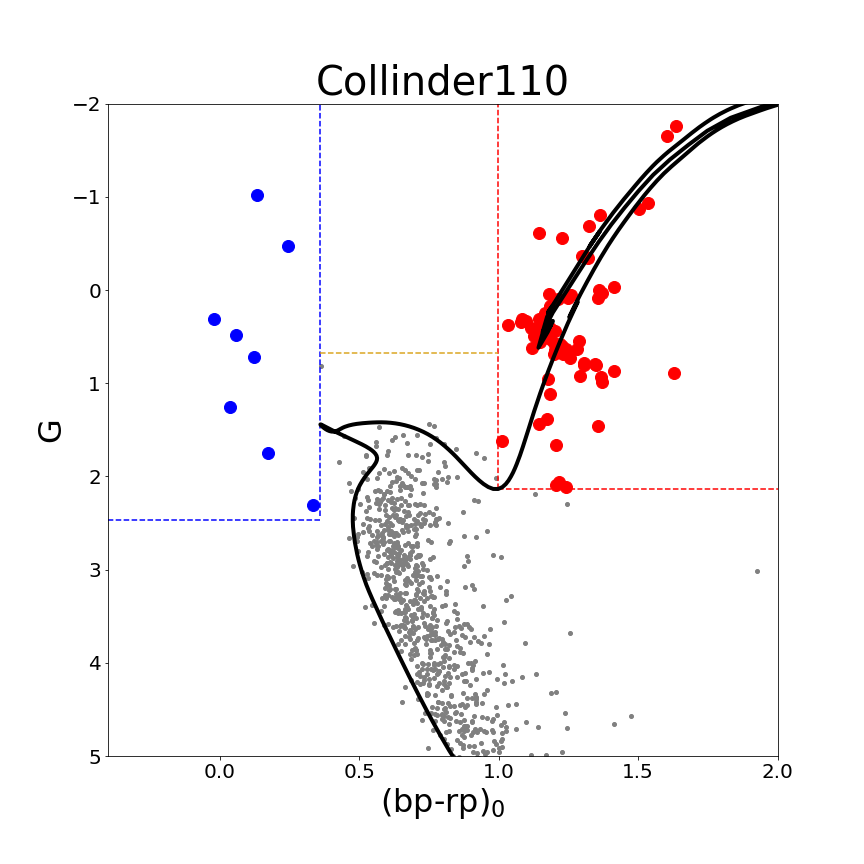}}
    \subfigure[]{\includegraphics[width=.3\linewidth]{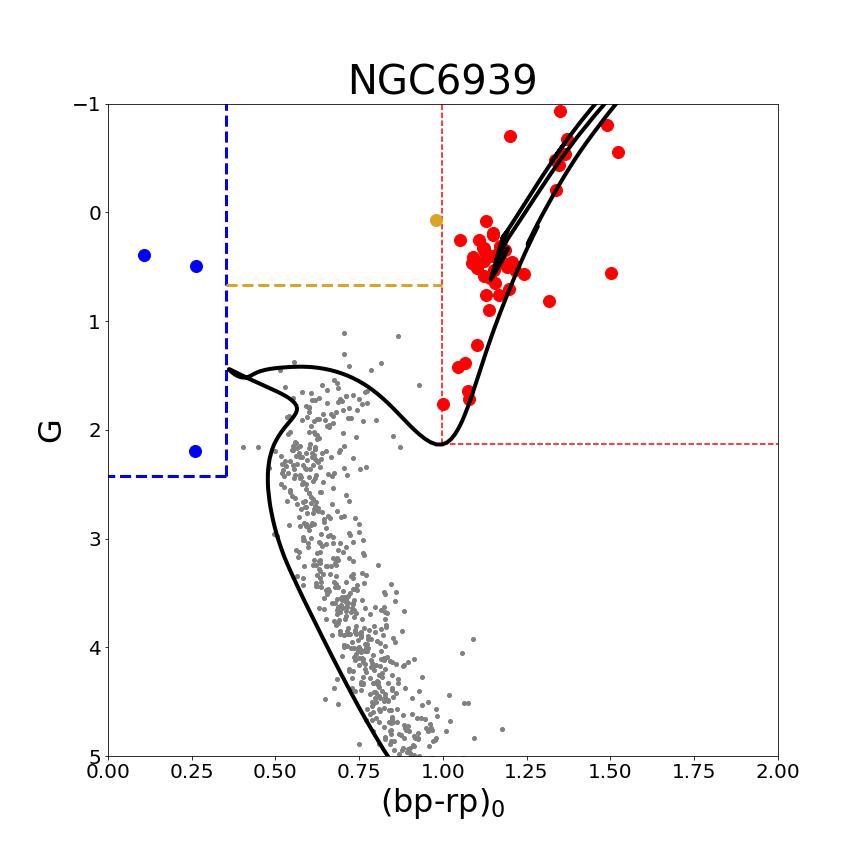}}
    \subfigure[]{\includegraphics[width=.3\linewidth]{new_cmds/IC465112pc_cmd.png}}
    \subfigure[]{\includegraphics[width=.3\linewidth]{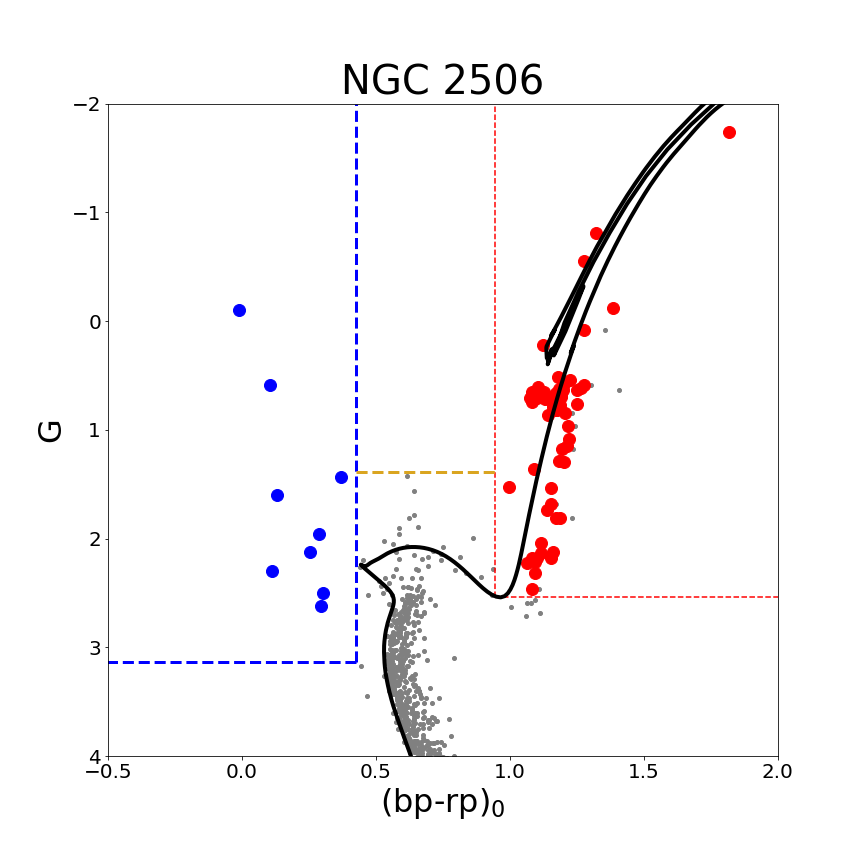}}
    \subfigure[]{\includegraphics[width=.3\linewidth]{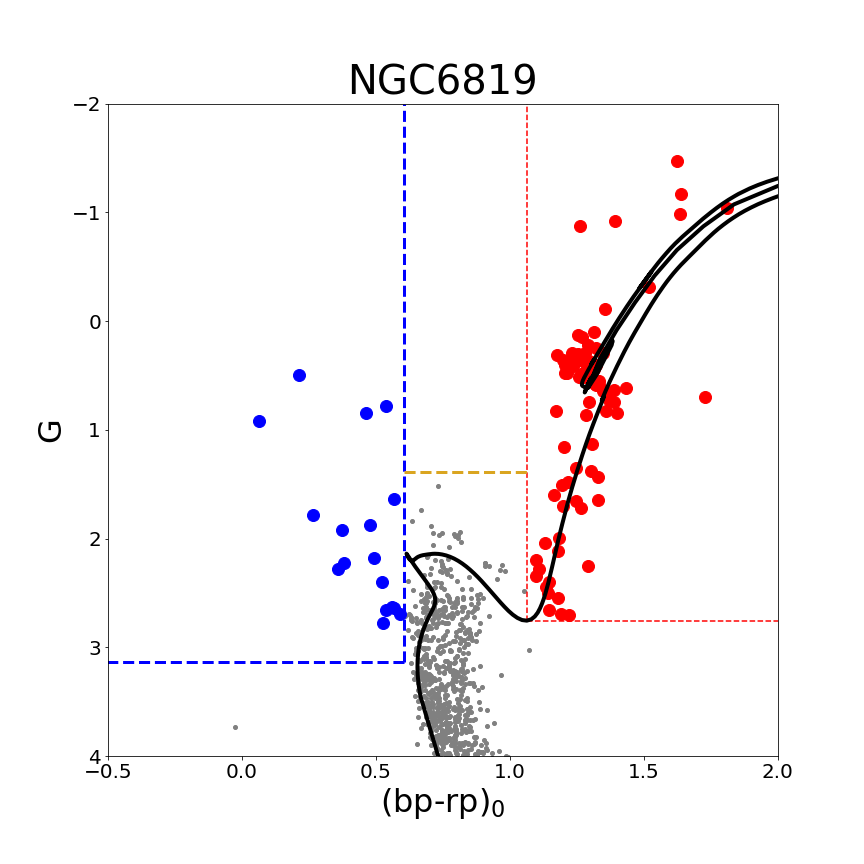}}
    \subfigure[]{\includegraphics[width=.3\linewidth]{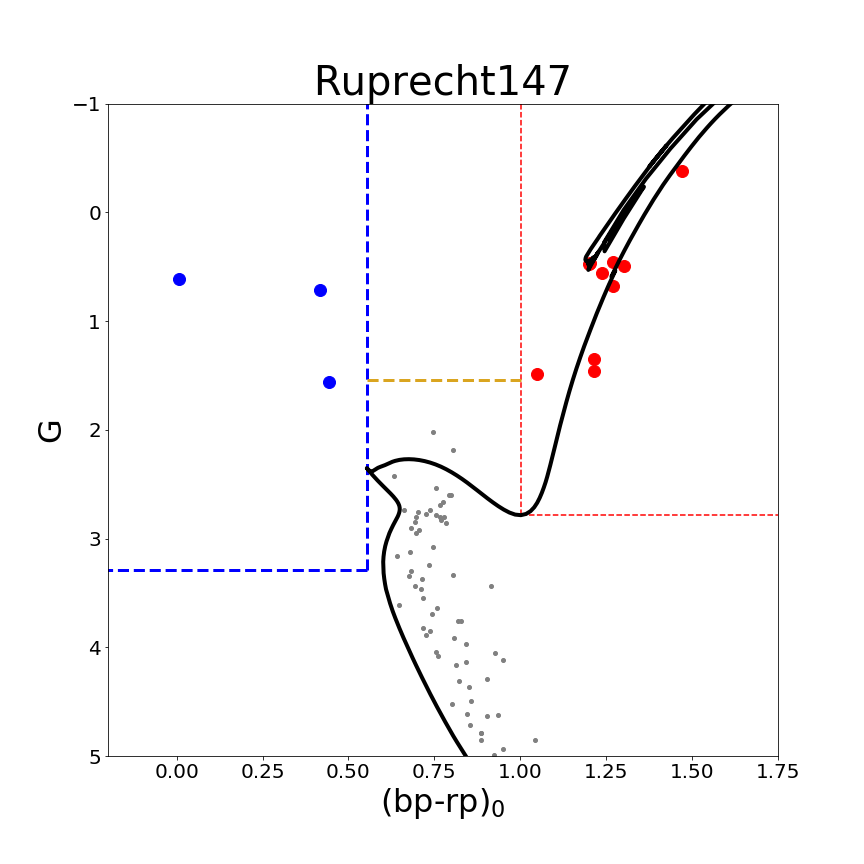}}
    \subfigure[]{\includegraphics[width=.3\linewidth]{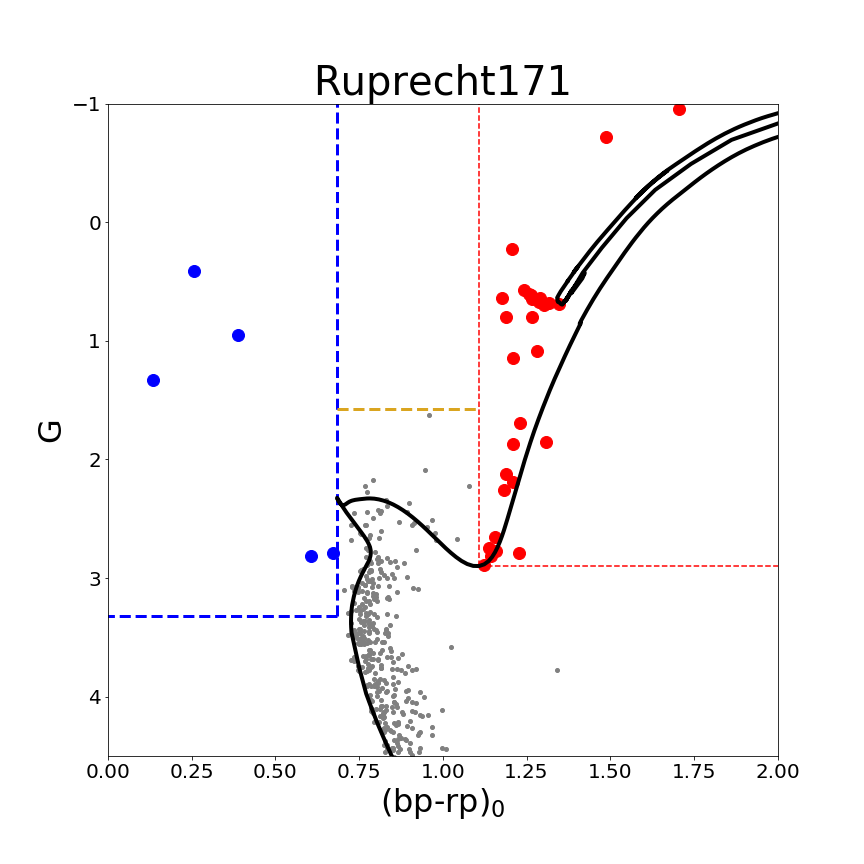}}
\end{figure*}

\begin{figure*}

    \subfigure[]{\includegraphics[width= .3\linewidth]{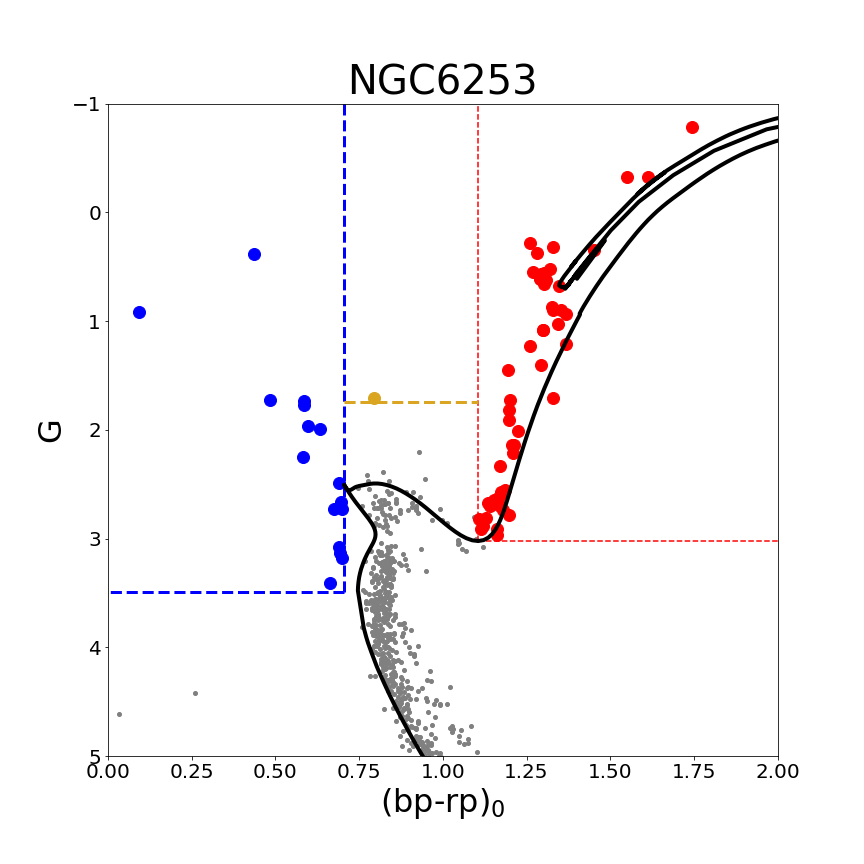}}
    \subfigure[]{\includegraphics[width=.3\linewidth]{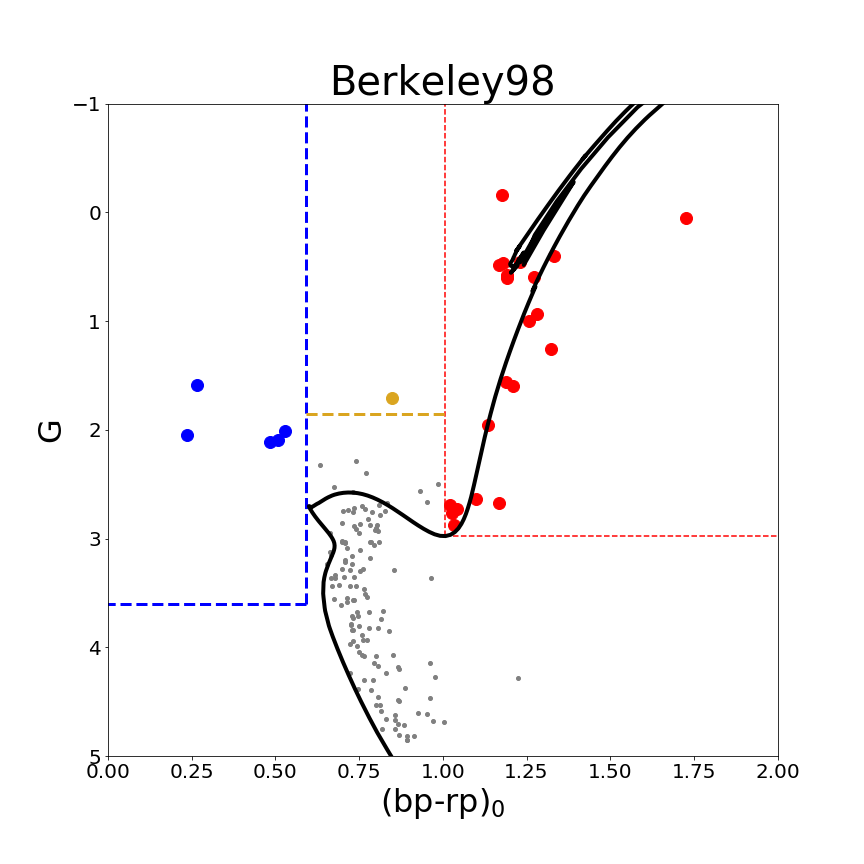}}
    \subfigure[]{\includegraphics[width= .3\linewidth]{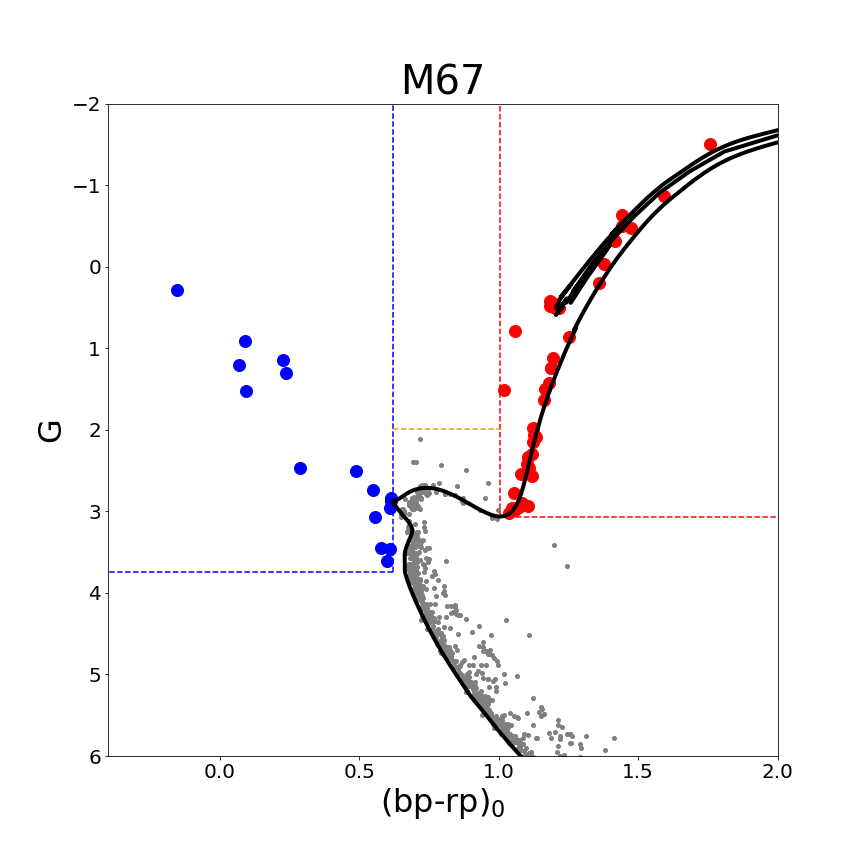}}
    \subfigure[]{\includegraphics[width=.3\linewidth]{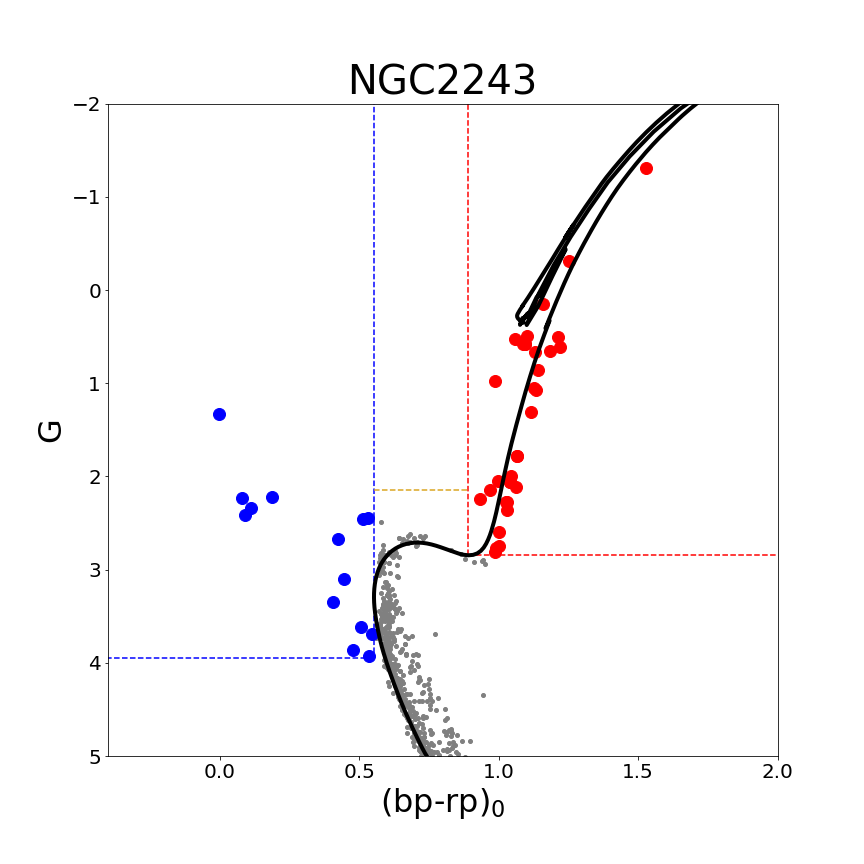}}
    \subfigure[]{\includegraphics[width=.3\linewidth]{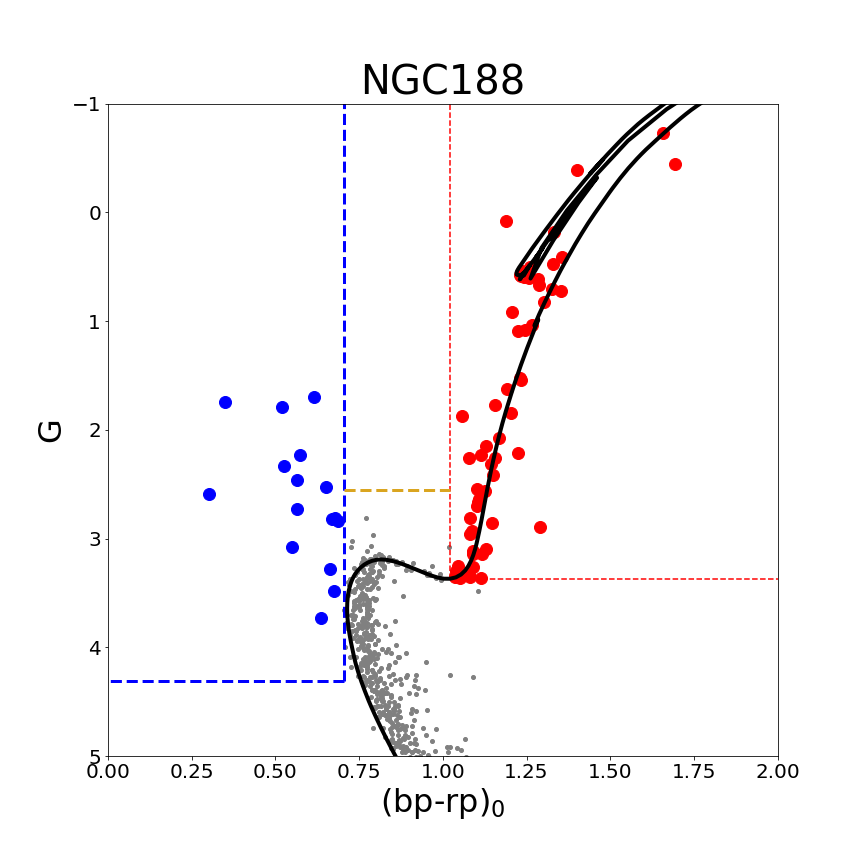}}
    \subfigure[]{\includegraphics[width=.3\linewidth]{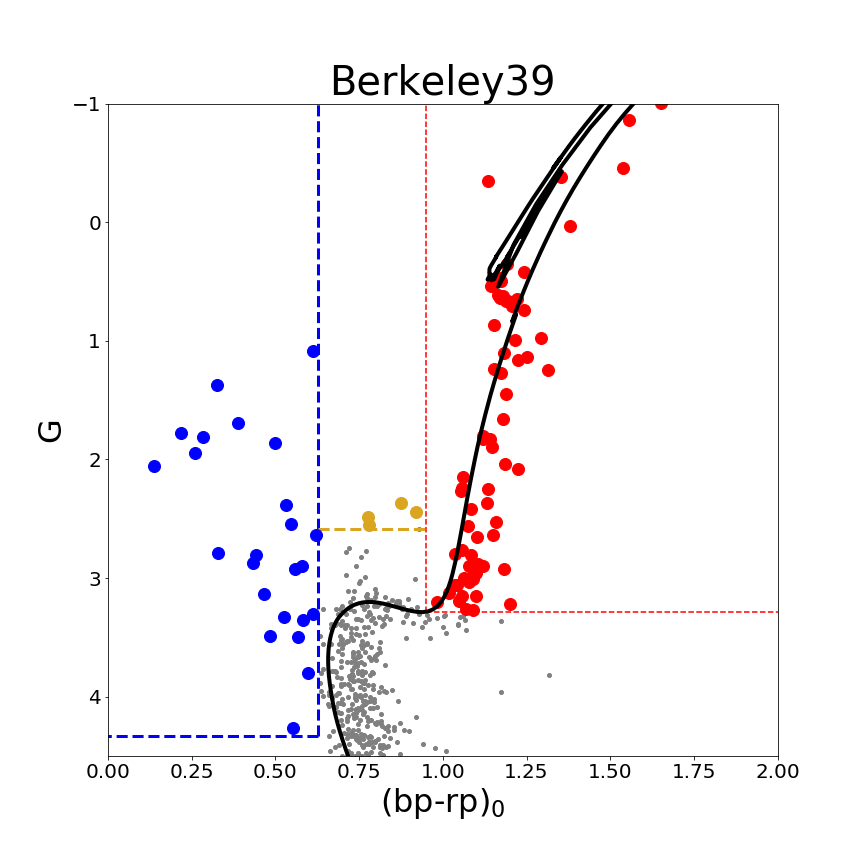}}
    \subfigure[]{\includegraphics[width= .3\linewidth]{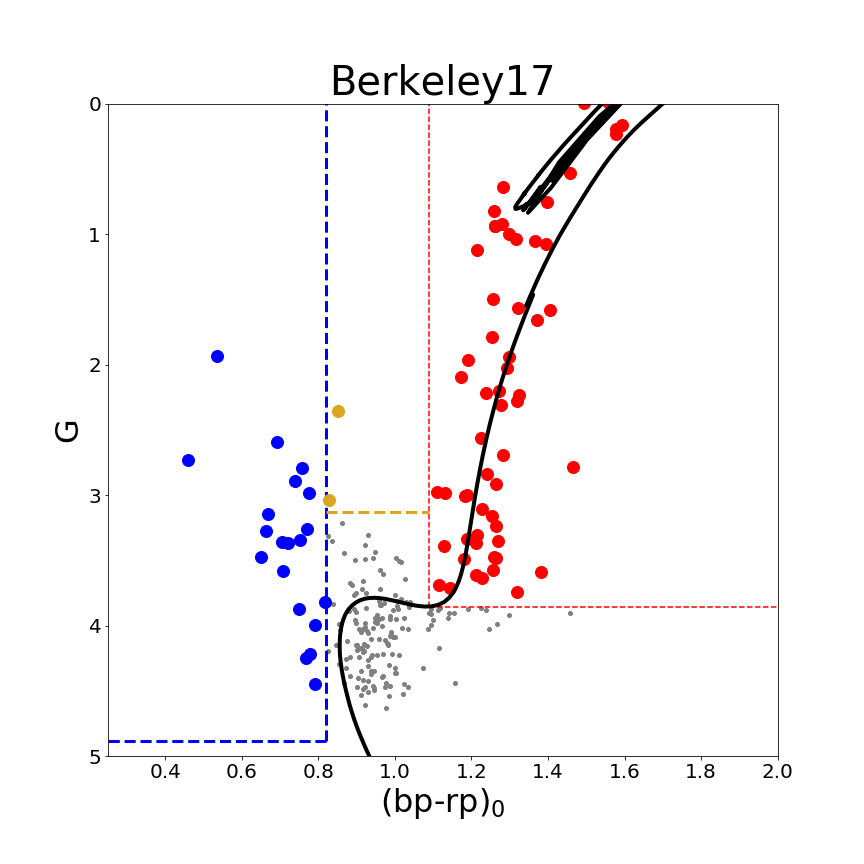}}
\centering
   \caption{Color-magnitude diagram showing Gaia members of the cluster (gray points). Colored points mark the RGB stars (red), blue stragglers (blue) and yellow straggler region (yellow). The blue straggler region is brighter and bluer than the blue dashed lines. The RGB region is redder and brighter than the red dashed lines. The yellow straggler region falls between the red and blue regions brighter than 0.75 magnitude above the turnoff (yellow line). We also show MIST \citep{Dotter2016} isochrones with the age and metallicity given in Table 1.}
    \label{fig:allclustercmds}
\end{figure*}

 \begin{figure*}
     \centering
    \includegraphics[width= .95\linewidth]{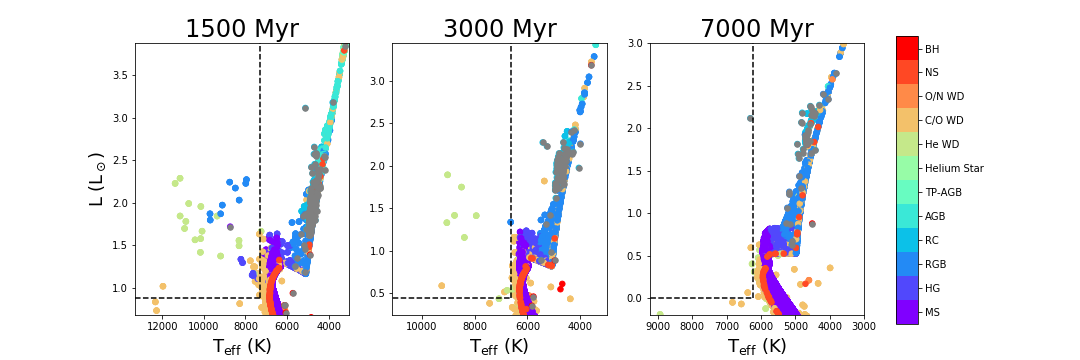}
    \includegraphics[width= .95\linewidth]{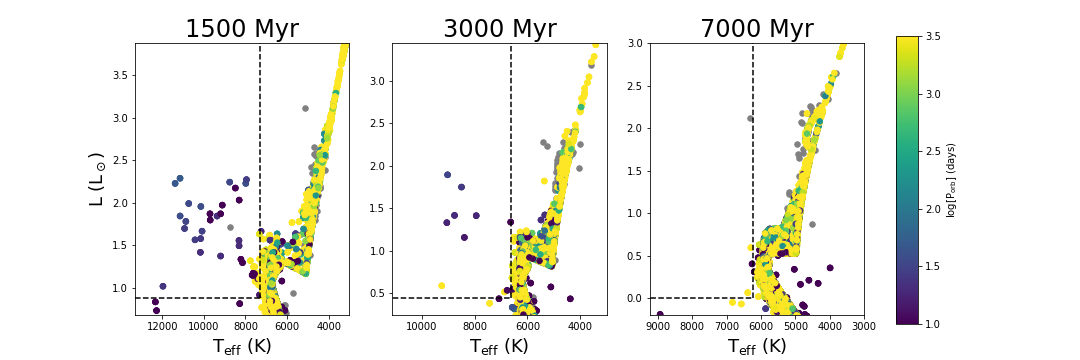}
     \caption{HR diagrams showing the results of COSMIC population synthesis simulations of 100,000 binaries using \citet{Hurley2002} mass transfer stability prescription. We show three snapshots in age: 1500 Myr (left), 3000 Myr (middle), and 7000 Myr (right). In the top panels we plot stars colored by the stellar type of the primary (i.e., the originally more massive star in our simulation). In the bottom panels we show the same simulations, but we color the stars by orbital period. Gray points indicate systems that are now single stars (i.e., due to a binary merger) and thus do not have orbital periods. The dashed lines indicate the boundaries of the blue straggler domain.}
     \label{fig:HTPcmd}
 \end{figure*}
 
\begin{figure*}
    \centering
    \includegraphics[width= .95\linewidth]{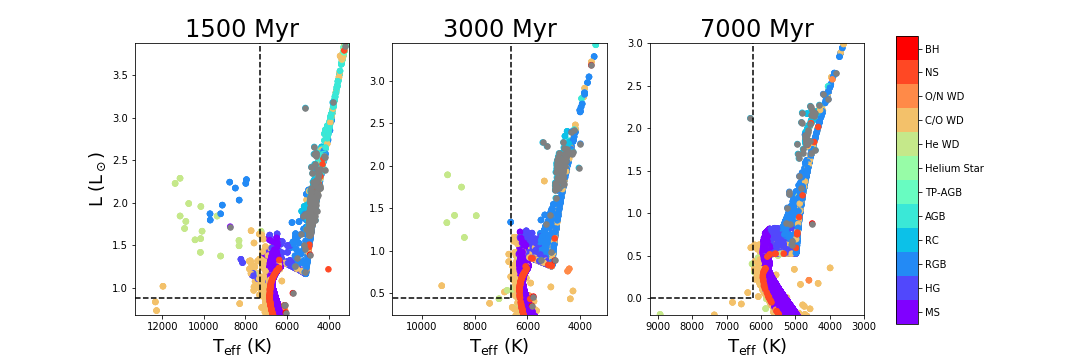}
    \includegraphics[width= .95\linewidth]{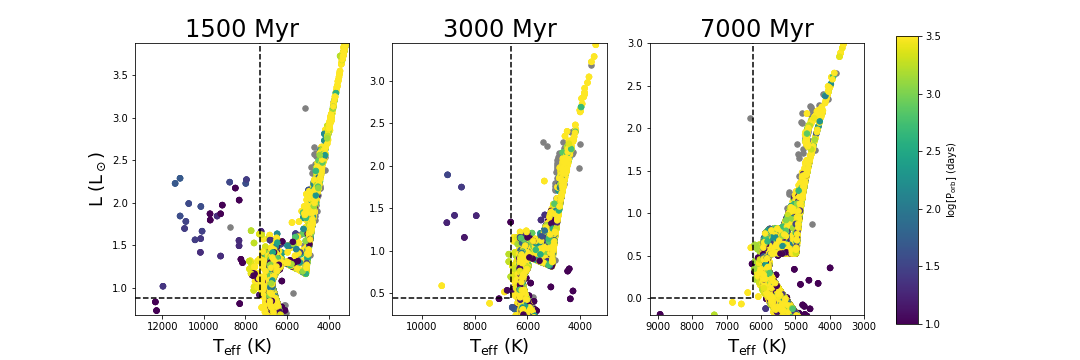}
    \caption{Same as Figure ~\ref{fig:HTPcmd}, but showing results using the \citet{Hjellming1987} prescription for mass transfer from RGB and AGB donors.}
    \label{fig:NWcmd}
\end{figure*}

 \begin{figure*}
     \centering
    \includegraphics[width= .95\linewidth]{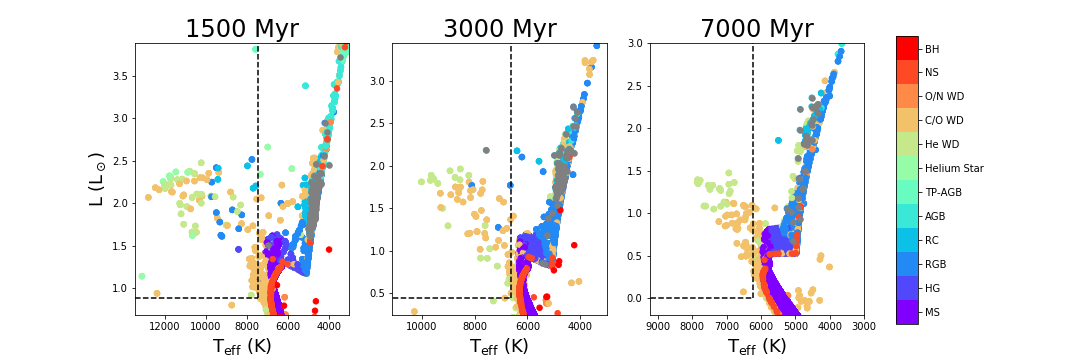}
    \includegraphics[width= .95\linewidth]{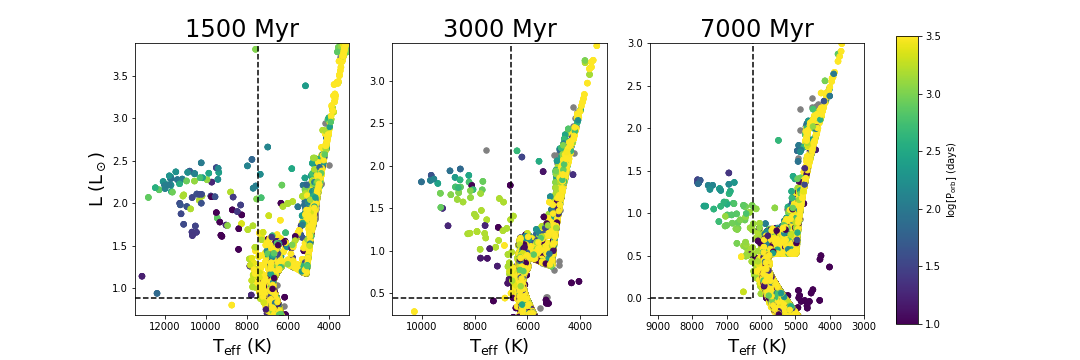}
     \caption{Same as Figure ~\ref{fig:HTPcmd}, but showing results using our non-conservative \qcrit\ prescription for mass transfer from RGB and AGB donors.}
     \label{fig:NCcmd}
 \end{figure*}
 
  \begin{figure*}
     \centering
    \includegraphics[width= .95\linewidth]{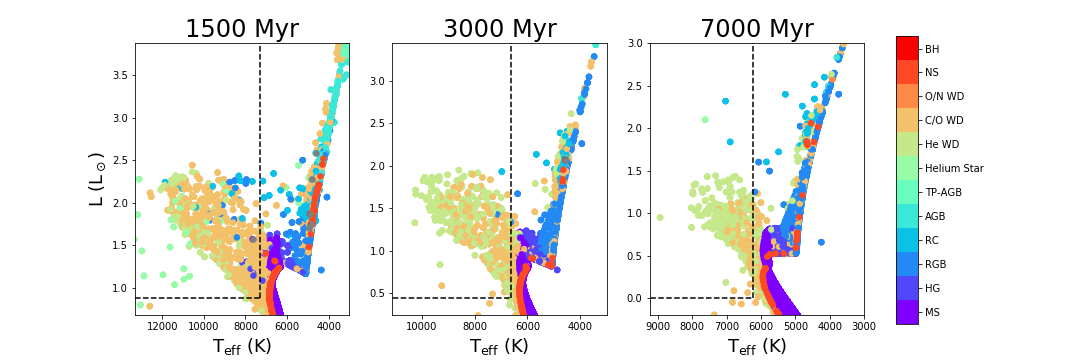}
    \includegraphics[width= .95\linewidth]{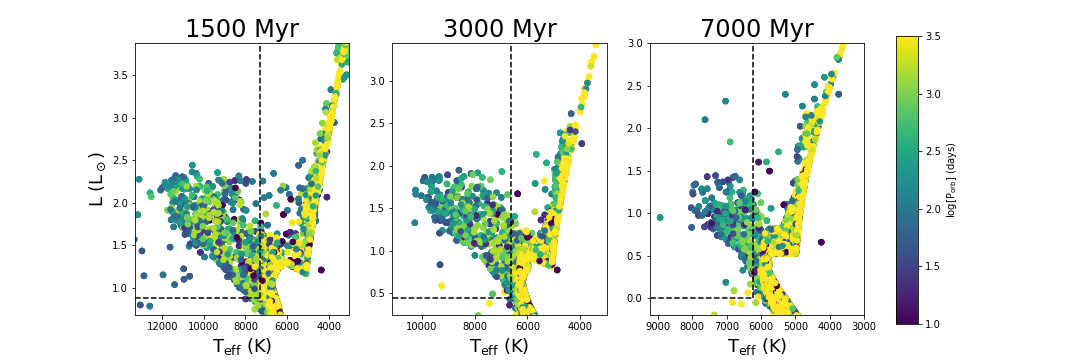}
     \caption{Same as Figure ~\ref{fig:HTPcmd}, but showing results using $q_\text{crit}= 100.$ for mass transfer from RGB and AGB donors, making all mass transfer stable.}
     \label{fig:NoCEHR}
 \end{figure*}

 \end{appendix}
 
\end{document}